\documentclass[journal]{IEEEtran}
\usepackage[T1]{fontenc}
\usepackage{textcomp}
\usepackage[utf8]{inputenc}
\usepackage{array}
\usepackage{multirow}
\usepackage{amsmath}
\usepackage{graphicx}
\usepackage{setspace}
\usepackage[bookmarks=true,bookmarksnumbered=true,bookmarksopen=true,bookmarksopenlevel=1,
 breaklinks=false,pdfborder={0 0 0},pdfborderstyle={},backref=false,colorlinks=false]
 {hyperref}
\hypersetup{pdftitle={Your Title},
 pdfauthor={Your Name},
 pdfpagelayout=OneColumn, pdfnewwindow=true, pdfstartview=XYZ, plainpages=false}

\makeatletter

\newcommand{\lyxmathsym}[1]{\ifmmode\begingroup\def\b@ld{bold}
  \text{\ifx\math@version\b@ld\bfseries\fi#1}\endgroup\else#1\fi}

\let\SF@@footnote\footnote
\def\footnote{\ifx\protect\@typeset@protect
    \expandafter\SF@@footnote
  \else
    \expandafter\SF@gobble@opt
  \fi
}
\expandafter\def\csname SF@gobble@opt \endcsname{\@ifnextchar[%]
  \SF@gobble@twobracket
  \@gobble
}
\edef\SF@gobble@opt{\noexpand\protect
  \expandafter\noexpand\csname SF@gobble@opt \endcsname}
\def\SF@gobble@twobracket[#1]#2{}
\providecommand{\tabularnewline}{\\}

\usepackage[caption=false,font=footnotesize]{subfig}
\usepackage{cite}
\usepackage{indentfirst}
\usepackage{hyperref}

\makeatother

\begin{document}
\title{Sustainable Air–Ground Integrated Coverage Networks: ISCC Architecture,
Technologies, and Testbed}
\author{Junyu~Liu,~\IEEEmembership{\textit{Member},~IEEE},~Xiayu~Zhang,~\IEEEmembership{Graduate~Student~Member,~IEEE}\textit{,}~Min~Sheng,~\IEEEmembership{Fellow,~IEEE},\\~Ruichen~Zhang\textit{,}~\IEEEmembership{Member,~IEEE}\textit{,}~Nan~
Zhao,~\IEEEmembership{\textit{Senior}~\textit{Member},~IEEE}\textit{,}~Jiacheng~Wang,~\IEEEmembership{Member,~IEEE},\\~and~Jiandong~Li,~\IEEEmembership{Fellow,~IEEE}\thanks{J. Liu, X. Zhang, M. Sheng, and J. Li are with the State Key Laboratory
of Integrated Service Networks, Institute of Information Science,
Xidian University, Xi'an 710071, China{\footnotesize{} (e-mail: junyuliu@xidian.edu.cn,
22011110207@stu.xidian.edu.cn, \{msheng, jdli\}@mail.xidian.edu.cn)}.
R. Zhang and J. Wang are with the College of Computing and Data Science,
Nanyang Technological University, Singapore (e-mail: \{ruichen.zhang,
jiacheng.wang\}@ntu.edu.sg). Nan Zhao is with the School of Information
and Communication Engineering, Dalian University of Technology, Dalian,
Liaoning 116024, China (e-mail: zhaonan@dlut.edu.cn). }}
\maketitle
\begin{abstract}
The rapid emergence of sixth-generation (6G) networks and the low-altitude
economy has accelerated the evolution of wireless infrastructures
toward air–ground integrated coverage networks (AGICNs), which seamlessly
fuse terrestrial and aerial communication resources. However, existing
AGICN studies primarily focus on coverage enhancement, while ignoring
sustainability. Pursuing sustainable AGICNs introduces new challenges
due to the multidimensional resource coupling across heterogeneous
air–ground segments. In view of this, this paper presents a comprehensive
survey and tutorial on sustainable AGICNs, aiming to balance coverage
capacity with carbon efficiency in low-altitude economies. An integrated
sensing, communication, and computation (ISCC)-driven architecture,
which enables dynamic resource orchestration through closed-loop control,
is proposed. We thus introduce a multi-dimensional sustainability
metric system, which covers operational efficiency, task-oriented
performance, and full lifecycle carbon emissions, to quantify energy
and carbon footprints. We review enabling technologies, including
artificial intelligence, hybrid precoding, integrated sensing and
communication, and simultaneous wireless information and power transfer,
and discuss their integration into the ISCC framework to minimize
energy consumption while maintaining robust coverage. Experimental
results on a real-world testbed demonstrate a 20\% reduction in power
consumption while achieving over 90\% coverage probability, highlighting
the feasibility of sustainable AGICNs for future green networks.
\end{abstract}

\begin{IEEEkeywords}
Air–ground integrated coverage networks (AGICNs), integrated sensing,
communication, and computation (ISCC), sixth-generation (6G), sustainability. 
\end{IEEEkeywords}

\section{Introduction}

\subsection{Background}

\IEEEPARstart{T}{he} sixth generation (6G) of wireless communication
systems is envisioned to deliver intelligent, ubiquitous, and sustainable
connectivity, enabling immersive services including holographic communications,
digital twins, and autonomous systems on a global scale. Against this
backdrop, the low-altitude economy is emerging as a new driver of
global economic growth \cite{EH_7,EH_8}. It encompasses a broad ecosystem
of low-altitude aerial platforms, including uncrewed aerial vehicles
(UAVs), urban air mobility systems, and high-altitude platforms (HAPs),
enabling applications that range from instant logistics and agricultural
operations to disaster response, environmental monitoring, and aerial
passenger transport \cite{Technology_1}. These emerging services
impose unprecedented demands on wireless communication systems. They
require ultra-reliable and ultra-low-latency control signaling, as
well as seamless coverage in complex three-dimensional (3D) urban
environments to support high-rate data transmission \cite{UAV_1,UAV_2}.

Traditional terrestrial cellular networks are not originally designed
to support aerial users. Their antenna beams are predominantly downward-tilted,
which leads to severe coverage gaps at altitude and introduces inter-mainlobe
interference \cite{Handover_4,History_3}. To address these limitations,
aerial–ground integrated coverage networks (AGICNs) have emerged as
a promising solution \cite{History_4}. AGICNs build upon terrestrial
cellular infrastructure, while incorporating aerial communication
nodes, including UAVs operating as aerial base stations (BSs) or relays
\cite{UAV_shortage_12,UAV_shortage_11}. This integration forms a
multi-layered and dynamically reconfigurable 3D heterogeneous network
architecture \cite{History_4}. The goal is to jointly exploit the
stability and high capacity of terrestrial infrastructure together
with the flexibility, broad coverage, and on-demand deployment capabilities
of aerial nodes, thereby overcoming the challenges associated with
low-altitude coverage \cite{Recent_study_3,ISCC_HAP_5}.

While AGICNs provide ubiquitous connectivity, they also introduce
extra energy consumption and carbon emissions, which pose a long-term
sustainability challenge \cite{Recent_study_4}. On the terrestrial
side, energy demand increases sharply. To support AGICNs, cellular
BSs must deploy massive multiple-input multiple-output (MIMO) arrays
and operate in high-frequency bands to overcome severe aerial–to-ground
(A2G) or ground–to-aerial (G2A) path loss and enable 3D beamforming
\cite{Channel_4}. This results in a significant increase in the power
consumption of radio frequency (RF) chains and signal processing modules
\cite{UAV_shortage_6}. On the aerial side, the energy profile is
fundamentally different. UAVs and other aerial platforms consume most
of their energy in propulsion systems to maintain hovering or movement
\cite{UAV_shortage_10}. This propulsion demand far exceeds the energy
required for communication functionalities \cite{UAV_shortage_9,UAV_shortage_12}.
Consequently, large-scale UAV deployments require frequent battery
charging or replacement, thereby increasing operational costs and
overall carbon emissions \cite{ISCC_HAP_5}. 

It is worth mentioning that, traditional network optimization frameworks
primarily emphasize energy efficiency (EE), defined as the number
of bits transmitted per joule \cite{UAV_Shortage_5}. However, high
EE does not necessarily imply low carbon emissions \cite{goverment2003energy}.
A system with excellent EE may still produce unacceptable carbon emissions
if its electricity is sourced from high–carbon-intensity fuels, e.g.,
coalfired electricity \cite{ministry2020green}. Consequently, the
sustainability targets of 6G require AGICNs to evolve from communication-centric
and EE-driven designs toward green, intelligent systems that pursue
carbon neutrality across their full operational lifecycle \cite{ISCC_HAP_4,UAV_shortage_3}.

The realization of sustainable AGICNs faces a series of theoretical
and practical challenges stemming from the multi-dimensional coupling
of communication, energy, and environmental dynamics \cite{UAV_shortage_9,UAV_shortage_10}.
The coexistence of terrestrial and aerial segments introduces highly
heterogeneous communication links characterized by varying altitudes,
mobility patterns, and propagation conditions \cite{Background_1,AGIN_4}.
Capturing these heterogeneous interactions within a unified analytical
framework is crucial yet challenging, particularly when incorporating
stochastic renewable energy models and carbon accounting mechanisms
\cite{Low-carbon-economy-REF1,Low_carbon_economy_1}. Furthermore,
as sustainable AGICNs expand to global-scale deployments, ensuring
energy and communication management across massive aerial and terrestrial
nodes demands hierarchical architectures, efficient signaling, and
carbon-efficient cooperation mechanisms \cite{Recent_study_3,Low-carbon-economy-REF8}.
The interplay between decentralized decision-making and global sustainability
objectives introduces fundamental coordination challenges that cannot
be solved by traditional network control models \cite{History_4}.
To address these challenges, the integrated sensing, communication
and computation (ISCC) paradigm has been introduced, advocating a
shift from isolated functional design toward deep cross-domain integration
\cite{ISCC_2,ISCC_1}. ISCC is widely regarded as a native capability
of future 6G networks. Building on this foundational paradigm, we
aim to develop a carbon-aware closed-loop intelligence framework that
unifies sensing, communication, and computation to achieve sustainable
AGICNs.

\begin{figure}[t]
\begin{centering}
\includegraphics[width=8.5cm]{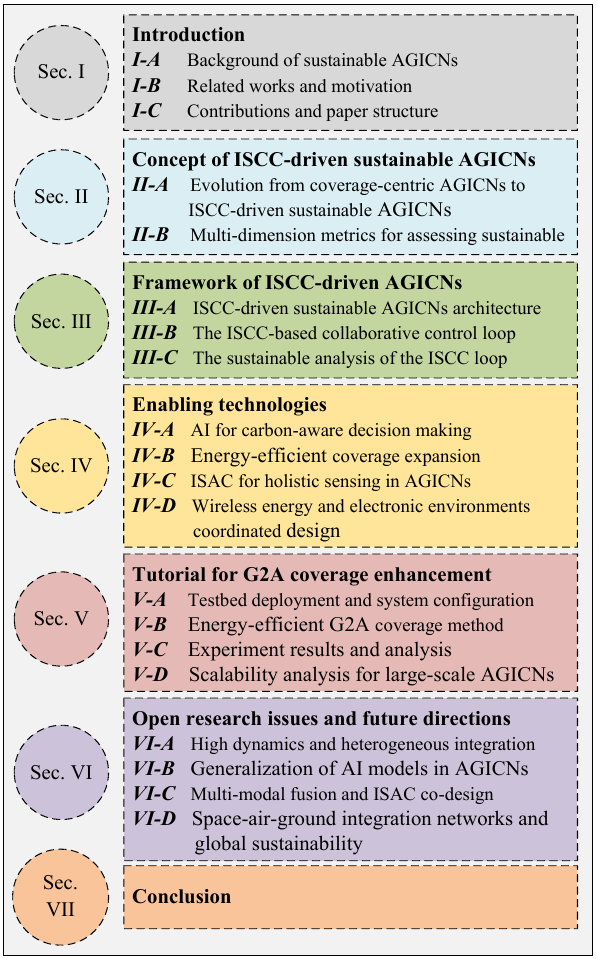}
\par\end{centering}
\caption{Outline and structure of this work. \label{fig:Outline-and-structure}}
\end{figure}

\subsection{Motivation and Contributions}

To achieve sustainable AGICNs, existing studies \cite{ISCC_HAP_5,SWIPT_5,SWIPT_10,AGIN_COM_6,Recent_study_5,Recent_study_3,Technology_1,Recent_Study_2,Recent_study_1,Recent_study_4}
have primarily focused on EE and energy consumption optimization,
identifying the limited onboard energy of UAVs as the fundamental
challenge to network sustainability. Minimizing energy consumption
and improving EE have become the core objectives for extending the
network’s operational endurance and lifecycle. Specifically, studies
\cite{ISCC_HAP_5,Recent_study_5,Recent_study_4} have shown that for
high-mobility UAVs, propulsion energy consumption accounts for the
dominant portion of total energy usage. Optimizing flight trajectory
and velocity to reduce propulsion energy is considered a more effective
approach than merely decreasing transmission power. Leveraging the
passive and low-power characteristics of reflecting intelligent surfaces
(RIS), studies \cite{Recent_Study_2,AGIN_COM_6,Technology_1} have
further explored the integration of RIS with UAVs to enhance A2G link
quality and EE. Furthermore, to prolong the lifetime of AGICNs, studies
\cite{Recent_study_1,SWIPT_5,SWIPT_10} have introduced renewable
energy sharing and wireless power transfer (WPT) to supply energy
to UAVs, thereby reducing dependence on traditional high-carbon power
grids. 

At the system level, the 6G vision calls for a more holistic framework
in which sensing, communication, computation, and energy interact
synergistically to enhance network intelligence, adaptability, and
carbon efficiency. Studies \cite{ISCC_2} and \cite{ISCC_intro_1}
established the conceptual framework of ISCC and emphasize the multi-dimensional
co-design of time–frequency, power, and computing resources to enhance
system performance. To reveal the information-theoretic limits of
ISCC systems, study \cite{ISCC_theory_1} characterized the trade-off
between signal-to-noise ratio (SINR) and sensing accuracy, providing
guidance for system design. In terms of sensing and computing integration,
\cite{ISCC_task_oriented_2} introduced an ISCC framework with over-the-air
computation, achieving up to 280\% relative improvement in inference
accuracy (from 25\% to 95\%) by maximizing the minimum pair-wise discriminant
gain. Moreover, green-oriented ISCC design has also gained significant
attention. \cite{ISCC_green_1} developed a full-duplex ISCC framework
for low-altitude networks, where coordinated beamforming and joint
sensing–communication resource allocation reduce redundant signaling.
This design improves spectral–energy efficiency by over 45\% and lowers
end-to-end energy consumption by 52\% compared with half-duplex ISCC
baselines. Complementarily, \cite{ISCC_green_2} introduced a UAV-enabled
ISCC architecture enhanced by split federated learning (FL), which
adaptively partitions training between UAVs and ground nodes. This
approach reduces onboard computation energy by 30\%, cuts communication
overhead by 28\%, and boosts inference accuracy by 12\% under limited
power budgets. Together, these works demonstrate the quantitative
sustainability gains achievable through learning-driven and coordination-aware
ISCC designs.

Despite these promising advances, existing ISCC studies mainly focused
on static or ground-based network settings, where the spatiotemporal
dynamics of aerial–terrestrial interactions are not fully captured.
Moreover, most approaches optimize energy consumption or computation
efficiency in isolation, without establishing a unified sustainability
control loop that accounts for carbon awareness, adaptive resource
coupling, and lifecycle energy impacts \cite{AGIN_1,Shortcome_3}.
In this work, building on the emerging ISCC paradigm, we investigate
how sensing, communication and computation can be jointly orchestrated
to enable sustainable operation of AGICNs. Our analysis shows that
achieving low-carbon and energy-efficient coverage is no longer merely
a physical-layer optimization problem. Instead, it increasingly relies
on artificial intelligence (AI)-driven sensing, prediction and decision
making distributed across aerial and terrestrial nodes. In particular,
we distinguish sustainable AGICNs, which embed carbon awareness and
energy intelligence into the ISCC control loop, from conventional
communication-centric AGICNs that prioritize coverage extension or
throughput maximization. With this distinction in place, we systematically
study how AI-enabled ISCC supporting carbon-aware state estimation,
adaptive resource coordination, multimodal energy sensing and cross-layer
optimization under diverse low-altitude service demands. These insights
form the core contributions of this work, which are summarized as
follows.
\begin{itemize}
\item This study first establishes a comprehensive framework for low-carbon
AGICNs from the perspectives of sustainability, intelligence, and
carbon efficiency. The fundamental characteristics of coverage-oriented,
EE-oriented, and ISCC-driven sustainable network architectures are
clarified. By establishing a multidimensional coupled analysis framework,
the study elucidates the intrinsic interdependencies among communication,
sensing, computation, and energy subsystems in achieving carbon-efficiency
optimization, providing a theoretical foundation for the green evolution
of 6G low-altitude infrastructure.
\item To address the limitations of conventional performance metrics, we
innovatively propose a triple-layer evaluation framework encompassing
operational efficiency, task effectiveness, and full life-cycle assessment.
The framework captures energy consumption during operation, and the
implicit carbon footprint associated with device manufacturing and
decommissioning, providing criteria for subsequent algorithm design,
network optimization, and experimental evaluation. Grounded in this
framework, this work systematically reviews and categorizes pathways
for achieving energy saving and carbon reduction in AGICNs across
four dimensions, including computing, communication, sensing, and
energy. 
\item A ISCC-based AI control loop for sustainable AGICNs is proposed and
detailed. Specifically, the sensing module acquires multi-source spatiotemporal
information, while the computation module, deployed at the edge or
in the cloud, executes optimizers based on deep reinforcement learning
(DRL) to generate control information. The communication module efficiently
delivers this information and enforces decisions to provide reliable
G2A coverage. As part of the tutorial component of this paper, a large-scale
field validation is carried out in Zigong, Sichuan, which verifies
the proposed framework and the effectiveness of the ISCC-based AI
control loop, providing practical guidance for future deployment.
\end{itemize}

\subsection{Paper Organization}

The paper is organized as follows (as shown in Fig. \ref{fig:Outline-and-structure}).
Section II establishes the foundational background, examining the
evolution from coverage-centric to sustainable AGICNs and introducing
key performance metrics for evaluating the network's coverage and
sustainability trade-off. Section III proposes the ISCC-driven framework,
detailing the synergistic integration of communication, sensing, and
computation subsystems into a closed-loop control architecture for
carbon-aware optimization. Section IV provides a comprehensive review
of enabling technologies, including AI-empowered decision-making,
energy-efficient beam management, Integrated sensing and communication
(ISAC), and Wireless energy harvesting (WEH), analyzing their roles
in reducing carbon emissions. Section V validates sustainable G2A
coverage through a testbed. Section VI discusses future directions.
Finally, Section VII concludes with the sustainable evolution path
for low-carbon economic infrastructures.

\section{Background and Motivation of Low-carbon AGICNs}

In this section, we first review the evolution of sustainable AGICNs,
including the coverage-centric design of the fourth and fifth generation
(4G and 5G) systems, the EE-oriented design introduced in 5G advanced,
and the ISCC-driven sustainable AGICNs envisioned for 6G. To quantitatively
assess the sustainability of AGICNs, we review a set of indicators,
including operational efficiency metrics, task-oriented metrics, and
full life-cycle metrics. 

\subsection{Evolution of Sustainable AGICNs}

The development of AGICNs has not been achieved overnight. Rather,
their design objectives and enabled technologies are undergoing a
profound paradigm shift. We categorize this evolution into three main
stages. Specifically, coverage-centric AGICNs, energy-efficient AGICNs,
and ISCC-Driven Sustainable AGICNs

\subsubsection{Coverage-Centric AGICNs}

In early 4G/5G development, AGICNs primarily aimed for seamless terrestrial-to-aerial
coverage and high downlink capacity from aerial BSs \cite{UAV_shortage_12}.
Researchers employed convex or heuristic methods to optimize deployment,
tilt, subchannels, and power allocation, without explicitly targeting
energy saving, only imposing constraints like transmission or propulsion
power limits \cite{EH_3,ISCC_HAP_5}. Consequently, these coverage-centric
designs incurred inherently high overall energy consumption.

\subsubsection{Energy-Efficient AGICNs}

In the mid-to-late 5G era, with the rise of the low-altitude economy,
researchers recognized the severe energy consumption in AGICNs, especially
UAV propulsion energy. Studies \cite{UAV_shortage_10,UAV_shortage_11}
investigate propulsion power models and shift the optimization objective
from maximizing coverage capability to improving EE. Much work focused
on joint trajectory and communication optimization, examining trade-offs
between flight speed and energy use \cite{Recent_study_5,UAV_ISAC_1}.
However, most research treated all energy as homogeneous regardless
of source, so AGICNs achieved high EE but still suffered from high
carbon emissions \cite{UAV_energy_study_3,UAV_energy_study_4}.

\subsubsection{ISCC-Driven Sustainable AGICNs}

In the 6G communication era, ISCC technology enables networks to integrate
sensing, communication, and computation functions deeply. This represents
a fundamental shift in design objectives, evolving from performance-oriented
goals in Stage 1 and efficiency-oriented goals in Stage 2 to green
and intelligent goals in Stage 3 \cite{UAV_energy_study_5}. At this
stage, the optimization objective focuses on minimizing total carbon
emissions or achieving carbon neutrality while satisfying users' quality
of service (QoS), sensing, and computation constraints \cite{UAV_energy_study_3}.
ISCC drives the transformation of AGICNs from passive energy saving
to proactive carbon-aware and carbon-efficient operation.

\subsection{Evaluation Metrics for Sustainable AGICNs }

The ISCC-driven sustainable AGICNs discussed in Section 2.1 necessitates
the establishment of a new set of multi-dimensional sustainability
indicators to comprehensively evaluate the performance of ISCC-driven
AGICNs. Depending on the task requirements supported by ISCC-driven
AGICNs, we categorize the sustainability metrics into three dimensions,
instantaneous metrics reflecting operational efficiency, task-oriented
performance metrics, and lifecycle-oriented metrics for capturing
long-term energy and carbon impacts. 

\subsubsection{Operational Efficiency Metrics}

These metrics aim to measure the instantaneous operational efficiency
of the AGICNs, including refined EE, carbon intensity, and renewable
energy ratio. Specifically, these metrics are stated as follows. 

\paragraph{Refined EE (REE, Eq. (\ref{eq:REE})). \textmd{On the basis of the
original definition of EE, the REE is defined as}}

\textit{
\begin{equation}
\mathrm{REE}=\frac{R_{\mathrm{total}}}{P_{\mathrm{total}}^{\mathrm{Ref}}}\,[\mathrm{bits/Joule}],\label{eq:REE}
\end{equation}
where $R_{\mathrm{total}}$ is system's total transmission bits and
$P_{\mathrm{total}}^{\mathrm{Ref}}=P_{\mathrm{Ground}}+\sum_{u\in\Pi_{\mathrm{UAV}}}(P_{\mathrm{Comm,u}}+P_{\mathrm{Prop,u}}+P_{\mathrm{Comp,u}})$.
$P_{\mathrm{Ground}}$, $P_{\mathrm{Comm,u}}$, $P_{\mathrm{Prop,u}}$
and $P_{\mathrm{Comp,u}}$ are ground BS's energy, UAV's communication
energy, UAV's propulsion energy, and computing energy, respectively.
In this work, we consider a radio access network (RAN)-level energy
model, where $P_{\mathrm{Ground}}$ includes the transmit power and
baseband processing power of the BS, while the energy consumption
of the core network and backhaul is not included. $\Pi_{\mathrm{UAV}}$
is UAV set \cite{UAV_energy_study_1}. }

\paragraph{Carbon Intensity (CI, Eq. (\ref{eq:CI})). \textmd{CI refers to }\textit{$\mathrm{CO}_{2}$}\textmd{
emissions generated by AGICNs when transmit per bit }\cite{CI-REF1}}

\textit{
\begin{equation}
\mathrm{CI}=\frac{C_{\mathrm{total}}}{R_{\mathrm{total}}}\,[\mathrm{g\cdot CO_{2}/bit}],\label{eq:CI}
\end{equation}
where $C_{\mathrm{total}}=C_{\mathrm{Ground}}+C_{\mathrm{UAVs}}$
is system's total $\mathrm{CO}_{2}$ emissions. $C_{\mathrm{Ground}}$
represents the carbon emissions associated with the ground communication
network, which is calculated based on the energy consumption of the
BS and the corresponding carbon intensity factor. Consistent with
the energy model, only the RAN-level emissions are considered, excluding
the contributions from the core network and backhaul. $C_{\mathrm{UAVs}}$
is UAVs' carbon emissions, which determined by carbon emissions of
battery charging and discharging. }

\paragraph{Renewable Energy Ratio (RER, Eq.(\ref{eq:RER})). RER is defined
as the ratio between the renewable energy consumption and the total
energy consumption \cite{UAV_energy_study_6}}

\begin{equation}
\mathrm{RER}=\frac{P_{\mathrm{renew}}}{P_{\mathrm{total}}}\,[\%],\label{eq:RER}
\end{equation}
\textit{which can be improved by dispatching UAVs to areas with sufficient
solar energy for photovoltaic charging.}

\subsubsection{Task-Oriented Metrics}

These indicators aim to measure the sustainability of ISCC when it
performs communication, sensing, and computation tasks.

\paragraph{Coverage Energy Efficiency (CEE, Eq.(\ref{eq:CEE})). CEE quantifies
the effective coverage area achieved per unit of energy consumption
\cite{CEE-REF1,CEE-REF2}}

\begin{align}
\mathrm{CE} & \mathrm{E}\label{eq:CEE}\\
= & \frac{\mathrm{Effective\:wireless\:coverage\:area}}{\mathrm{Transmit\:energy}}\,[\mathrm{\mathrm{km}^{2}\,or\,\mathrm{km}^{3}/Joule}],\nonumber 
\end{align}
\textit{which indicates the coverage capability that can be provided
with the lowest energy consumption. When measuring A2G's CEE, the
unit for the effective wireless coverage area should be $\mathrm{km}^{2}$.
While the unit is $\mathrm{km}^{3}$, when measuring G2A's CEE.}

\paragraph{Sensing Energy Efficiency (SEE, Eq.(\ref{eq:SEE})). SEE measures
the amount of environmental information collected per joule of energy
consumed \cite{SEE-REF1,SEE-REF2}}

\begin{equation}
\mathrm{SEE}=\frac{\mathrm{\mathrm{CRB}^{-1}}}{\mathrm{Sensing\:energy}}\,[\mathrm{bits/Joule}],\label{eq:SEE}
\end{equation}
\textsl{where cramér-rao bound (CRB) is the sensing error and its
inverse means the Fisher information for unbiased estimators. SEE
represents the efficiency to obtain high-quality sensing information,
including positioning accuracy, environmental monitoring with the
lowest energy consumption.}

\paragraph{Computational Efficiency (CompE, Eq.(\ref{eq:CE})). CompE evaluates
the efficiency of computing, especially when AGICNs are performing
AI reasoning and data analysis tasks \cite{CE-REF1}}

\begin{align}
\mathrm{Com} & \mathrm{pE}\label{eq:CE}\\
= & \frac{\mathrm{Floatingpoint\:operations\:per\:second}}{\mathrm{Computation\:energy}}\,[\mathrm{FLOPs/Joule}],\nonumber 
\end{align}
\textsl{which can be utilized to decide whether the computing task
is performed on the UAV or offloaded to the ground.}

\subsubsection{Full Life-Cycle Metrics}

Full life-cycle metric plays a pivotal role in quantifying the total
carbon emissions generated throughout the entire life span of AGICNs,
from raw material extraction and manufacturing to transportation,
operation, maintenance, and end-of-life disposal.

\paragraph{Full Life Cycle Carbon Emissions (FLCCE, Eq.(\ref{eq:FLCCE})) \cite{FLCCE-REF1}}

\begin{align}
\mathrm{FLCCE}\:(\mathrm{kg\cdot CO_{2}})\nonumber \\
=C_{\mathrm{prod}}+C_{\mathrm{trans}} & +C_{\mathrm{oper}}+C_{\mathrm{maint}}+C_{\mathrm{dispos}},\label{eq:FLCCE}
\end{align}
\textit{where $C_{\mathrm{prod}}$ denotes emissions from the production
phase, including material extraction and equipment manufacturing.
$C_{\mathrm{trans}}$ accounts for emissions from transportation and
logistics. $C_{\mathrm{oper}}$ captures emissions during the operational
phase, typically the dominant contributor. $C_{\mathrm{maint}}$ represents
emissions from maintenance and component replacement. $C_{\mathrm{dispos}}$
corresponds to the emissions during disposal, recycling, or decommissioning. }

\paragraph{Embodied Carbon Return Period (ECRP, Eq.(\ref{eq:ECRP})). ECRP represents
the time period to save the $C_{\mathrm{oper}}$ realized through
ISCC to offset the embodied carbon generated by manufacturing equipment\protect\footnote{https://www.oecd.org/content/dam/oecd/en/publications/reports/2021/02/ready-for-take-off\_64c92c83/6e3f6792-en.pdf}}

\begin{equation}
\mathrm{ECRP}=\frac{C_{Embodied}}{\mathrm{Operational\:carbon\:saving\:per\:year}}\,[\mathrm{years}],\label{eq:ECRP}
\end{equation}
\textit{where $C_{Embodied}=C_{\mathrm{prod}}+C_{\mathrm{trans}}+C_{\mathrm{maint}}$
is the carbon emissions caused by AGICNs equipment production, transportation,
and maintenance.}

These metrics constitute the foundation of a multi-dimensional, task-oriented
ISCC-driven AGICN, enabling fine-grained assessment and targeted optimization
of system-level energy consumption and carbon emissions, and thereby
driving the transition from traditional capacity-centric design toward
carbon-aware, energy-efficient networking paradigms.

\begin{figure*}[t]
\centering
\includegraphics[width=14cm]{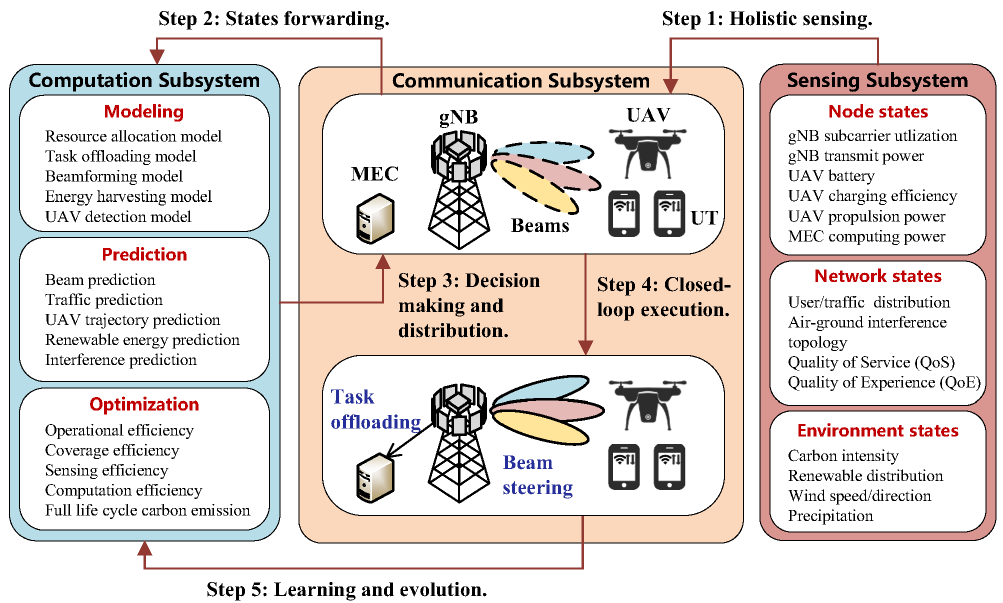}

\caption{The ISCC-based collaborative control loop for sustainable AGICNs.
\label{fig:The-ISCC-based-collaborative}}
\end{figure*}

\section{The Framework of ISCC-Driven Sustainable AGICNs}

In Section II, we establish the evolutionary objectives of sustainable
AGICNs. In this section, we explore how these objectives can be achieved
at the system level through deconstructing the internal architecture
and functional coupling mechanisms of the ISCC-driven AGICNs paradigm.
We further elaborate on how ISCC serves as the intelligent module
of AGICNs, enabling carbon-aware closed-loop control.

\subsection{ISCC-Driven AGICNs Architecture}

The ISCC-driven sustainable AGICNs architecture depends on the close
collaboration among the three subsystems, i.e., communication subsystem,
sensing subsystem, and computation subsystem. These three subsystems
together constitute the \textquotedbl sensing-cognition-action\textquotedbl{}
ability of the AGICNs. Specifically, the three subsystems are detailed
as follows. 

\textbf{Sensing subsystem:} The sensing subsystem is responsible for
acquiring multidimensional and dynamic state information from heterogeneous
sources, forming the foundation of all intelligent decision-making
processes. Its sensing scope extends beyond traditional channel state
information and encompasses node states, network states, and environmental
states \cite{ISCC_1,ISCC_4}. The node state includes not only the
communication physical-layer parameters of UAV or BS, but also sustainability-related
attributes that reflect UAV’s operational condition, e.g., UAV’s onboard
battery, charging and discharging efficiency, and propulsion power
consumption \cite{UAV_energy_study_6,Recent_study_4}. The network
state captures information that reflects the overall system dynamics,
including user distribution, G2A interference topology, and user's
QoS, etc \cite{UAV_relay_2,Channel_2}. The environmental state includes
meteorological information closely related to UAV propulsion energy
consumption, e.g., wind speed, wind direction, and precipitation.
These environmental states are helpful to assist the computation subsystem
in planning UAV trajectories and sensing tasks \cite{UAV_energy_study_4,UAV_energy_study_3}.

\textbf{Communication subsystem:} In ISCC-driven AGICNs, the communication
subsystem not only provides seamless coverage and high-throughput
transmission to users, but also plays a critical role in maintaining
efficient and reliable information flow across the sensing subsystem
and computation subsystem \cite{ISCC_4,UAV_ISAC_6}. Specifically,
the communication subsystem supports transferring sensing data from
the sensing subsystem to the computation subsystem, exchanging state
information among nodes within the sensing subsystem, and delivering
control signals from the computation subsystem to the nodes, e.g.,
UAVs and BSs. These control signals include trajectory updates for
UAVs, antenna tilt adjustments for ground BSs, and task offloading
instructions for edge computing nodes, etc \cite{UAV_sensing_2,ISCC_UAV_1}.
It is worthwhile that the energy consumption and carbon emissions
caused by information flow transmission are non-negligible in ISCC-driven
AGICNs. 

\textbf{Computation subsystem:} The computation subsystem is responsible
for integrating massive amounts of sensing data, constructing various
task models, and performing real-time optimization methods. Its core
functions include modeling, prediction, optimization, and decision
making, etc \cite{ISCC_green_2}. Specifically, modeling refers to
the construction of a digital twin model based on historical and real-time
sensing data, which supports subsequent inference, prediction, and
optimization \cite{ISCC_5}. Prediction involves the exploit of AI
to estimate future states of system key parameters, including user
distribution, UAV energy consumption, and grid carbon intensity, thereby
providing essential inputs for decision making \cite{ISAC_beamforming_3}.
Optimization and decision making refer to the application of advanced
AI algorithms, including multi-agent DRL and FL, to solve complex
optimization problems in AGICNs, ensuring user's QoS while enhancing
system’s overall sustainability \cite{ISCC_8,ISCC_7}. 

\subsection{The ISCC Control Loop for Sustainable AGICNs}

In Section 3.1, we reviewed the functions and operating mechanisms
of each subsystem in ISCC. In this section, we discuss how these subsystems
collaborate to form a closed-loop control architecture that enables
sustainable AGICNs (as shown in Fig. \ref{fig:The-ISCC-based-collaborative}).
This closed loop consists of five components: holistic sensing, state
forwarding, decision making, closed-loop execution, learning and evolution.

\textbf{Step 1: Holistic sensing \cite{ISCC_loop_1}. }The sensing
subsystem continuously collects node, network, and environmental states.
Typical sensing information includes UAV battery levels, user distribution,
and the carbon intensity of the power grid. In practice, certain variables,
such as future traffic demand, environmental dynamics, and carbon
intensity, may be only partially observable. In such cases, the sensing
module can be complemented by estimation or prediction mechanisms
based on historical data or learning-based models. The closed-loop
control structure further mitigates the impact of imperfect observations
by continuously updating decisions based on newly available information
\cite{AI_2,AI_6}.

\textbf{Step 2: State forwarding \cite{ISCC_loop_2}. }The communication
subsystem transmits the sensed information to the computation subsystem.
For example, UAVs send their battery information to edge computing
nodes through uplink connections, or act as relays to forward data
collected from ground user terminals to the computation subsystem.
\begin{table*}[tbh]
\caption{Energy Consumption of ISCC Components. \label{tab:Energy-Consumption-of}}

\centering{}%
\begin{tabular}{|c|c|c|c|}
\hline 
{\scriptsize\textbf{ISCC Function}} & {\scriptsize\textbf{Component / Implementation}} & {\scriptsize\textbf{Power Consumption}} & {\scriptsize\textbf{Energy Characteristic}}\tabularnewline
\hline 
\multirow{4}{*}{{\scriptsize\textbf{Sensing \cite{ISCC_energy_1,ISCC_energy_2,ISCC_energy_3}}}} & {\scriptsize GPS} & {\scriptsize 0.05–0.5 W} & {\scriptsize Low-power, continuous operation}\tabularnewline
\cline{2-4}
 & {\scriptsize Visual camera} & {\scriptsize 1–5 W} & {\scriptsize Moderate, continuous perception}\tabularnewline
\cline{2-4}
 & {\scriptsize mmWave radar} & {\scriptsize 1–5 W} & {\scriptsize All-weather sensing, lower cost than LiDAR}\tabularnewline
\cline{2-4}
 & {\scriptsize 3D LiDAR} & {\scriptsize 10–50 W} & {\scriptsize High sensing accuracy, energy-intensive}\tabularnewline
\hline 
\multirow{4}{*}{{\scriptsize\textbf{Reasoning / Inference \cite{ISCC_energy_4,ISCC_energy_6}}}} & {\scriptsize TinyML / Neuromorphic} & {\scriptsize 0.01–0.1 W} & {\scriptsize Ultra-low-power inference}\tabularnewline
\cline{2-4}
 & {\scriptsize Embedded AI chip (Edge SoC)} & {\scriptsize 1–10 W} & {\scriptsize Lightweight real-time inference}\tabularnewline
\cline{2-4}
 & {\scriptsize Jetson Nano / Orin Nano} & {\scriptsize 1–10 W} & {\scriptsize Parallel multi-task inference}\tabularnewline
\cline{2-4}
 & {\scriptsize Jetson AGX Xavier} & {\scriptsize 10–100 W} & {\scriptsize High-performance decision-making}\tabularnewline
\hline 
\multirow{4}{*}{{\scriptsize\textbf{Training / Learning \cite{ISCC_energy_5,ISCC_energy_8}}}} & {\scriptsize On-device fine-tuning} & {\scriptsize 1–10 W} & {\scriptsize Local updates, low frequency}\tabularnewline
\cline{2-4}
 & {\scriptsize Federated learning (per agent)} & {\scriptsize 1–10 W} & {\scriptsize Distributed, accumulative cost}\tabularnewline
\cline{2-4}
 & {\scriptsize Edge server training} & {\scriptsize 10–500 W} & {\scriptsize Offloaded computation}\tabularnewline
\cline{2-4}
 & {\scriptsize Centralized cloud training} & {\scriptsize 10–100 kW} & {\scriptsize High energy, intermittent}\tabularnewline
\hline 
\multirow{3}{*}{{\scriptsize\textbf{Action execution \cite{ISCC_energy_7,ISCC_energy_9}}}} & {\scriptsize Beamforming / phase control} & {\scriptsize < 1 W} & {\scriptsize Low-power electronic control}\tabularnewline
\cline{2-4}
 & {\scriptsize Transmit power (Micro/Pico BS)} & {\scriptsize < 10 W} & {\scriptsize Low-power communication}\tabularnewline
\cline{2-4}
 & {\scriptsize Transmit power (Macro BS)} & {\scriptsize < 1000 W} & {\scriptsize Dominant continuous energy consumption}\tabularnewline
\hline 
\end{tabular}
\end{table*}

\textbf{Step 3: Decision making \cite{UAV_energy_study_3,ISCC_loop_1}.}
The computation subsystem, composed of distributed AI agents, integrates
sensing information and employs modeling, prediction, and optimization
functions to generate decisions. The computation subsystem is driven
by operational efficiency metrics and task-oriented metrics, which
serve as optimization objectives or constraints, while full life-cycle
metrics are used to evaluate the overall sustainability performance
of the system.

\textbf{Step 4: Closed-loop execution \cite{ISCC_8,ISCC_6}. }The
communication subsystem disseminates the computed decisions to the
execution modules. When the communication subsystem serves as an execution
module, it can adjust network topology, antenna tilt angles, or transmission
power according to the received control signals. When computation
nodes act as execution modules, they can perform task offloading,
model training, or parameter updates, etc.

\textbf{Step 5: Learning and evolution \cite{ISCC_loop_3}.} To evaluate
the effectiveness of the executed actions, the sensing subsystem provides
feedback to the AI models within the computation subsystem in the
form of rewards or penalties. This feedback enables the models to
update and improve in subsequent control cycles, ensuring generalization
and real-time adaptability while preventing model degradation over
time.

The steps from 1 to 5 involve heterogeneous state variables and functional
modules operating on different time scales. Fast-varying states, such
as user distribution, interference topology, and channel conditions,
require frequent updates to support real-time control. In contrast,
slowly varying states, including grid carbon intensity and environmental
factors, evolve over longer time horizons and can be updated less
frequently. Accordingly, the ISCC framework adopts a time-scale separation
design. The sensing, decision-making, and execution modules form a
fast control loop operating at short time intervals, while the learning
and evolution module operates on a slower time scale, where model
updates are performed periodically rather than at every decision epoch.
This design decouples learning from real-time control, thereby improving
system stability and robustness.

\subsection{The Sustainable Analysis of the ISCC Control Loop}

While the ISCC framework enables significant energy and carbon reduction
of AGICNs, it also introduces additional energy consumption due to
sensing, communication, computation, and learning processes. To provide
a comprehensive understanding of the energy consumption introduced
by the ISCC framework, we summarize the typical power consumption
of different components, including sensing, reasoning/inference, training/learning,
and action execution, as shown in Tab. \ref{tab:Energy-Consumption-of}.

Tab. \ref{tab:Energy-Consumption-of} indicates the energy consumption
of ISCC components varying significantly across different modules.
Sensing and control-related operations typically incur low to moderate
energy consumption, while edge inference remains within a manageable
range. Although training processes, especially centralized cloud training,
are energy-intensive, they are performed intermittently rather than
continuously. In contrast, communication-related energy consumption,
particularly transmission power, is continuously incurred and often
dominates the overall energy budget. Therefore, by enabling more efficient
communication and reducing transmission power, the ISCC framework
can offset the additional energy overhead introduced by sensing, computation,
and learning, leading to an overall reduction in system energy consumption.

\section{Key Technologies for Sustainable AGICNs }

This section provides an in-depth discussion of the enabling technologies
required to realize the ISCC-driven AGICNs framework. The analysis
is conducted from four complementary dimensions, namely AI, energy-efficient
coverage expansion, ISAC, and SWIPT. We elaborate on how ISCC drives
and optimizes the technologies at each dimension and further examine
how these technologies, in turn, contribute to the implementation
of the ISCC control loop.

\subsection{AI for Carbon-Aware Decision Making}

The sustainable operation of AGICNs represents an inherently complex
optimization problem characterized by a high-dimensional state space,
which incorporates parameters including position, velocity, transmission
power, channel conditions, and battery energy. This problem further
involves non-convex objective functions and requires decision-making
within millisecond-level latency \cite{Recent_study_3}. Traditional
optimization techniques, e.g., convex optimization, are unable to
effectively address these challenges due to the strong parameter coupling,
non-convexity, and stringent real-time requirements \cite{ISCC_HAP_12}.
As a result, integrating AI into the computation subsystem becomes
a key enabler for achieving autonomous carbon reduction and intelligent
optimization in ISCC-driven AGICNs. Here, we summarize three major
learning technologies for AGICNs, i.e., reinforcement learning (RL),
supervised learning, and generative learning, along with their corresponding
applications. 

\textbf{RL in AGICNs. }Studies \cite{AI_1,AI_2} model UAV or BS as
a DRL agent wherein state information is generated by the ISCC sensing
subsystem. The state incorporates UAV battery energy, traffic distribution,
and other relevant parameters. The agent’s action space commonly includes
transmission power control, trajectory planning, and computation offloading
decisions. The reward function is formulated to reflect sustainability-oriented
objectives, including improvements in EE and carbon efficiency. For
instance, \cite{AI_1} develops a coverage-oriented path planning
method that employs an environmental map as the input to a convolutional
neural network and uses a double deep Q-network (DDQN) to learn UAV
control policies. The agent aims to maximize the covered area while
satisfying strict battery constraints, enabling adaptive planning
under highly dynamic environments and fluctuating energy levels. Another
representative work \cite{AI_2} designs an energy-saving path planning
algorithm in turbulent environments. This method incorporates diverse
flight modes including takeoff, landing, and horizontal flight, and
refines the reward function to minimize propulsion energy consumption.
Experimental evaluation shows that the planner based on RL reduces
the execution steps of UAV by more than 50\%. 

\textbf{Supervised learning and deep learning in AGICNs.} Supervised
learning and deep learning are widely employed in AGICNs for predictive
tasks, including mobility prediction, wireless channel estimation,
and traffic forecasting \cite{AI_3,AI_4,AI_5,AI_6}. These techniques
automatically extract temporal and spatial dependencies, thereby providing
intelligent support for network-level decision making. Leveraging
convolutional and autoencoder-based neural networks, \cite{AI_3}
proposes a two-stage denoising and least-squares estimation method
to estimate G2A wireless channel, which can reduce the bit error rate
of an 802.11ac orthogonal frequency division multiplexing system by
40–50\%. Deep learning models can also forecast network states and
traffic distributions, enabling proactive routing, scheduling, and
load balancing based on learned spatiotemporal patterns \cite{AI_4}.
By learning spatiotemporal traffic patterns, the system can anticipate
congestion or idle regions and accordingly adjust node activation
and load distribution. These predictive capabilities are essential
for low-carbon communication optimization, as accurate mobility and
traffic forecasts allow proactive adjustments of UAV or BS deployment
and transmission power, reducing unnecessary coverage and avoiding
resource waste, ultimately lowering energy consumption. 

\begin{figure*}[t]
\centering
\begin{centering}
\includegraphics[width=16cm]{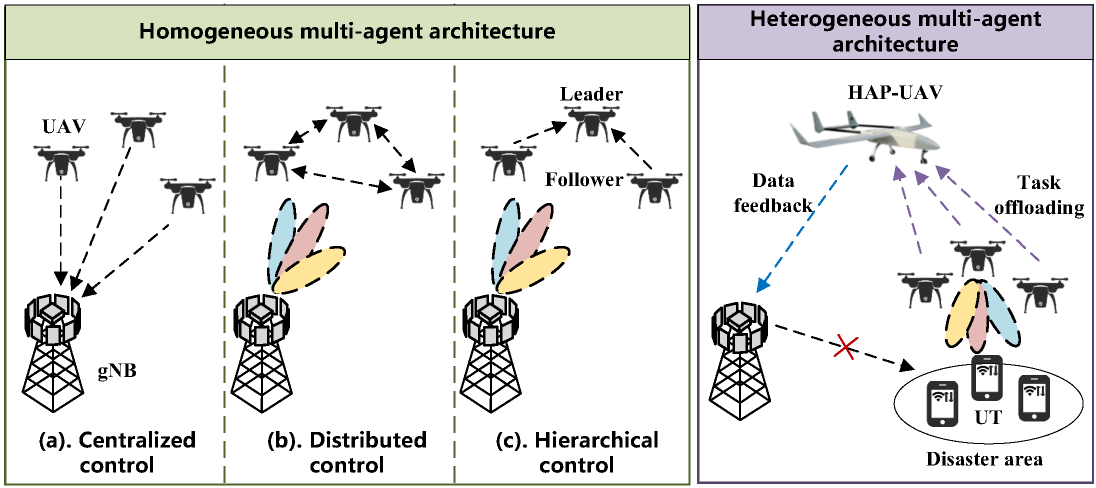}
\par\end{centering}
\caption{ISCC-enabled multi-UAV control architecture. \label{fig:ISCC-enabled-multi-UAV-control}}
\end{figure*}

\textbf{Generative model in AGICNs. }Generative models are capable
of producing new samples that approximate complex high-dimensional
distributions. Within AGICNs, these models primarily serve two purposes,
i.e., training enhancement and decision making \cite{AI_7}. For training
enhancement, generative models can synthesize diverse environmental
conditions, traffic patterns, and user distributions, which improves
the robustness of reinforcement learning–based control policies. For
decision making, generative models can learn the intrinsic structure
of network optimization problems and generate feasible solutions such
as UAV flight trajectories or resource allocation strategies \cite{AI_8}.
Existing studies \cite{AI_11,AI_9} show that generative models can
dynamically produce energy-efficient trajectories and resource allocation
strategies, improving both energy utilization and response efficiency.
Although generative approaches show strong potential in spatial decision-making
and scheduling, their effectiveness depends heavily on high-quality
training data and substantial computational resources \cite{AI_10}.
Furthermore, diffusion model has shown its potential in solving network
optimization problems and improving RL training efficiency, particularly
in high-dimensional and complex scenarios.

The studies mentioned above typically model the computation subsystem
as a single AI agent, while treating other subsystems or network nodes
in AGICNs as modules that execute the agent’s policy. However, as
the scale and complexity of AGICNs continue to grow, relying on centralized
AI computation can no longer meet future requirements for performance,
latency, and robustness. Therefore, pushing computational intelligence
to different nodes within the network and enabling distributed intelligence
is becoming an important trend, which corresponds to multi-agent collaboration.
In this work, two multi-agent architectures are investigated \cite{ISCC_UAV_swarm_1,ISCC_UAV_swarm_2}.
The first is a UAV swarm architecture, which consists of homogeneous
nodes working collaboratively to perform sensing, communication, and
computing tasks \cite{ISCC_UAV_swarm_3}. The second is a hierarchical
HAP-UAV architecture, where heterogeneous nodes at different altitudes
form a layered structure, combining the wide-area coverage and stability
of HAPs with the agility and flexibility of UAVs \cite{ISCC_UAV_swarm_4,ISCC_UAV_swarm_5}.
This section focuses on exploring these two architectures, highlighting
their unique advantages and application scenarios. 

\textbf{Homogeneous multi-agent architecture.} The UAV swarm architecture
is also known as a flying ad hoc network (FANET), which can be regarded
as the aerial extension of mobile and vehicular ad hoc networks \cite{ISCC_FANET_1,ISCC_FANET_3,ISCC_FANET_2}.
FANETs are characterized by 3D mobility, highly dynamic topology,
and uncertain connectivity, which lead to conventional routing protocols,
resource allocation strategies, and antenna or signal processing techniques
no longer applicable \cite{ISCC_FANET_1,ISCC_FANET_5,ISCC_FANET_4}.
To address these challenges, recent studies \cite{ISCC_FANET_6,ISCC_FANET_8,ISCC_FANET_7}
have explored various control architectures for UAV swarms, including
centralized control, distributed control, hierarchical control, and
bio-inspired control, as shown in Fig. \ref{fig:ISCC-enabled-multi-UAV-control}. 
\begin{itemize}
\item \textbf{Centralized control architecture.} In a centralized control
architecture, all decisions for the swarm are orchestrated by a single
central entity, e.g., a ground control station or a leader UAV. This
central node is responsible for collecting global information, computing
optimal strategies, and issuing commands to every member of the swarm
\cite{ISCC_FANET_9,ISCC_FANET_7}. However, this architecture may
suffer from single-point failure risk and poor scalability.
\item \textbf{Distributed control architecture.} To enhance resilience of
the UAV ad-hoc networks, a distributed control architecture delegates
decision-making authority to individual UAVs. Each UAV autonomously
makes decisions based on local observations and peer-to-peer communication
with neighboring nodes, without dependence on a centralized controller.
This distributed design significantly improves scalability, robustness,
and fault tolerance, especially when UAV networks suffer highly dynamic
mobility, link fluctuations, node failures, interference attacks,
and energy constraints \cite{ISCC_FANET_10,ISCC_FANET_11}.
\item \textbf{Hierarchical control architecture.} Hierarchical control architecture
is designed to organize the swarm into different layers or clusters,
resulting in a leader–follower structure \cite{ISCC_FANET_12,ISCC_FANET_14,ISCC_FANET_13}.
The leaders are responsible for high-level strategic planning, while
followers focus on tactical execution. This approach retains part
of the coordination benefits of centralized control while distributing
execution burdens, thereby achieving better scalability and real-time
responsiveness compared to a purely centralized model \cite{ISCC_FANET_15,ISCC_FANET_16}.
\item \textbf{Bio-inspired control architecture.} Furthermore, bio-inspired
control architectures have gained significant attention in recent
years. These approaches draw inspiration from the collective behaviors
of social organisms in nature, including ant colonies, bee swarms,
bird flocks, and fish schools \cite{ISCC_FANET_17,ISCC_FANET_19,ISCC_FANET_18}.
In such systems, complex global patterns emerge from simple, locally
governed interactions among individuals, without the need for any
central controller, providing a promising paradigm for scalable and
adaptive UAV swarm coordination.
\end{itemize}

\textbf{Heterogeneous multi-agent architecture.} The HAP–UAV hierarchical
architecture combines wide-area HAP coverage with the flexibility
of low-altitude UAVs \cite{AGIN_1,ISCC_HAP_1}. HAPs serve as access
points, backhaul hubs, computing centers, and macro-level coordinators,
while UAVs handle data collection, local processing, wireless access,
and latency-sensitive tasks \cite{ISCC_HAP_2,ISCC_HAP_4,ISCC_HAP_5}.
This architecture supports low-carbon sustainable networking in disaster
management, emergency communications, intelligent transportation,
and large-scale internet-of-things (IoTs) \cite{ISCC_HAP_6,ISCC_HAP_7}.
For instance, in disaster scenarios where ground infrastructure is
destroyed, HAPs act as aerial backhaul nodes connecting affected areas
to external control centers. At the same time, UAV swarms perform
real-time reconnaissance and act as mobile hotspots for survivors
and rescue teams. When UAVs face excessive loads, they offload tasks
to HAPs, preventing congestion and ensuring low-latency service delivery.

This subsection reviewed three AI paradigms for sustainable AGICNs,
specifically, RL for real-time control under dynamic states, supervised
learning for mobility and traffic prediction, and generative models
for scenario synthesis and direct solution generation. Multi-agent
architectures, including UAV swarms and HAP-UAV hierarchies, are highlighted
to address scalability. A common insight is that AI-based methods
outperform traditional optimization by adapting to non-stationary
environments, yet their performance remains sensitive to training
data quality and state observability.

\subsection{Energy-Efficient Coverage Expansion-Oriented AGICNs}

The communication subsystem, serving as the medium for information
exchange in ISCC-driven AGICNs, forms the foundation for achieving
energy savings and carbon reduction across the entire system. In AGICNs,
two fundamental communication modes exist depending on the role of
UAVs \cite{AGIN_4,AGIN_2}. The first mode involves cellular-connected
UAVs, where UAVs serve as users of ground cellular networks to enable
vertical industries, including passenger transport, parcel delivery,
and emergency medical services. The second mode refers to UAV-assisted
wireless communications, in which UAVs act as aerial communication
platforms equipped with transceivers to enhance network coverage,
particularly in applications including disaster recovery, relay transmission,
and coverage enhancement.

From the perspective of low-carbon communication, AGICNs offer high
flexibility, favorable channel conditions, and 3D signal propagation
capabilities, which pose two key challenges. First, when UAVs serve
as cellular users, their high altitude, uncertain trajectories, and
the exponential attenuation of radio signals with distance make it
particularly difficult to design energy-efficient and wide-coverage
G2A beams \cite{ISAC_beamforming_2}. Second, due to the favorable
G2A propagation conditions, adjacent G2A links often experience pronounced
mutual interference, which degrades the overall EE of the system \cite{Channel_1,Channel_2}.
In this work, we summarize energy-efficient strategies for AGICNs
from the perspective of beam management and interference cancellation. 

\textbf{Beamforming.} Beamforming is a technique that utilizes antenna
arrays to concentrate radiated energy in specific directions. This
concentration, known as beamforming gain, significantly enhances the
received signal strength at the target receiver without increasing
the total transmit power, thereby directly improving link EE \cite{BF_1,BF_2}.
In AGICNs, the ability to steer beams in 3D space (azimuth and elevation)
is critical. Leveraging the short wavelength characteristics of millimeter-wave
and higher-frequency bands, both aerial and terrestrial communication
platforms can integrate large-scale antenna arrays within limited
physical space, enabling precise beam alignment between air-ground
platforms and flexible adjustment of coverage areas as needed \cite{BF_3,BF_4}. 

\textbf{Hybrid beamforming architecture. }Although beamforming can
achieve highly energy-efficient communication through precise directional
control, its implementation typically relies on fully digital architecture
equipped with large-scale antenna. In fully digital architecture,
each antenna element is equipped with a dedicated RF chain, which
ensures the precise amplitude and phase control. Since the number
of RF chains equals the number of antennas, the computational complexity
of the baseband precoder is on the order of $\mathcal{O}(N_{\mathrm{T}}^{2})$,
where $N_{\mathrm{T}}$ is the number of transmit antenna \cite{BF_1,BF_5}.
Consequently, it results in extremely high energy consumption. Hybrid
precoding is a beamforming architecture that reduces hardware power
consumption by decreasing the number of RF chains through the coordinated
operation of analog and digital precoding layers. Study \cite{Comm_1}
employs deep neural networks to map compressed channel vectors to
analog beamforming vectors, enabling direct output of the precoding
matrix and effectively reducing training latency and energy overhead.
Study \cite{Comm_2} further introduces DRL into the hybrid precoding
framework, constructing an angle-domain-driven model that dynamically
selects active RF chains. This approach lowers channel estimation
complexity, and reduces energy consumption by more than 45\%. 

\textbf{Beam prediction}. AI-enabled fast beam prediction has become
a key pathway for reducing energy consumption and advancing low-carbon
communication \cite{Comm_5}. Traditional beam management relies on
frequent beam sweeping and high-precision channel estimation, which
incurs substantial computational overhead and energy expenditure \cite{Comm_6}.
AI models can extract spatial features from user locations and partial
channel measurements, and directly learn the nonlinear mapping between
these inputs and optimal beam directions. Study \cite{Comm_3} indicates
that Transformer models leverage the self-attention mechanism to capture
long-range dependencies in A2G channels, allowing the network to infer
future channel variations rather than relying solely on instantaneous
measurements. Experimental results show that a top-1 accuracy of 70\%
is obtained, and more than 35\% overhead is reduced. 

\textbf{Interference cancellation.} Recent studies demonstrate that
multi-agent reinforcement learning is effective for interference mitigation
and energy-efficient control in UAV-enabled networks. Studies \cite{Comm_7,Comm_8}
model UAVs as cooperative agents to jointly optimize trajectories
and transmit power, enabling coordinated interference mitigation.
Study \cite{Comm_10} proposes personalized anti-interference strategies
for heterogeneous UAV networks, allowing agents to select channels
and adjust power under adversarial interference via a federated SAC
method. Complementarily, study \cite{Comm_9} develops a hierarchical
multi-agent DRL framework in which agents handle dynamic power allocation
during cluster reconfiguration, supporting joint power reduction and
interference coordination. 

\textbf{Triangulation-based coordinated multiple points transmission
(CoMP).} Beyond physical-layer approaches, the CoMP capability of
terrestrial BSs can be exploited to suppress line-of-sight (LoS) interference
in G2A links, thereby reducing transmit power and energy consumption.
Study \cite{Testbed_2} proposes a Poisson–Delaunay triangulation-based
CoMP framework for G2A coverage in AGICNs. By optimizing the G2A coverage
topology, terrestrial BSs with highly correlated channels are grouped
into triangular clusters. Within each cluster, the three BSs jointly
form a triangular-prism coverage region through coordinated beam filling
in 3D space. Aerial users located in this coverage region are cooperatively
served by the three BSs. Numerical results in \cite{Triangular_1}
demonstrate that, in G2A coverage scenarios, the triangular-prism
structure reduces the spatial coverage overlap by more than 50\% compared
with conventional hexagonal terrestrial cellular structure. 

This subsection reviewed representative techniques for energy-efficient
coverage expansion in AGICNs, including advanced communication strategies
(e.g., hybrid beamforming, CoMP, and interference management), UAV-assisted
coverage enhancement, and wireless power transfer–enabled architectures
such as SWIPT and RIS/STAR-RIS. These approaches collectively aim
to improve spatial coverage while reducing transmission power and
enhancing energy utilization efficiency. A key insight is that, compared
with conventional static configurations, these methods achieve superior
performance by exploiting spatial diversity, cooperative transmission,
and adaptive deployment of aerial nodes.

\subsection{ISAC for Holistic Sensing in ISCC}

ISAC refers to the joint design and integration of sensing and communication
functionalities within a unified system, primarily focusing on the
efficient sharing of hardware, spectrum, and signal processing resources,
which introduces further opportunities for improving the sustainability
of AGICNs. At the hardware level, the joint use of radio-frequency
front ends, antenna arrays, and signal processing units can reduce
the deployment cost of physical equipment, i.e., $C_{\mathrm{prod}}$
in eq. (\ref{eq:FLCCE}). At the spectrum and waveform levels, the
ISAC functions decrease the total energy required for operations,
enhance both spectrum and energy efficiency, thereby mitigating the
operational carbon emissions, i.e., $C_{\mathrm{oper}}$ in eq. (\ref{eq:FLCCE}).
Building upon ISAC, ISCC further incorporates computing capabilities,
enabling not only the acquisition and transmission of information
but also its real-time processing, analysis, and decision-making.
Therefore, ISCC represents a more comprehensive paradigm that supports
a closed sensing–decision–action loop. Here, we summarize several
representative studies that investigate low-altitude ISAC scenarios
in AGICNs. 

\textbf{UAV acts as a sensing target or a communication node.} In
the early stages of the low-altitude economy, terrestrial cellular
network serves a dual function. By sharing hardware resources and
reusing time–frequency blocks, terrestrial BSs enhance UAV sensing
capability and ground-user communication performance under stringent
energy constraints. Study \cite{UAV_ISAC_4} investigates multi-BS
cooperative sensing and ISAC beam design to jointly optimize sensing
accuracy and communication performance under energy constraints. Building
on this idea, subsequent works \cite{ISCC_1,UAV_ISAC_6} extend UAVs
from passive sensing targets to integrated communication-and-sensing
nodes by jointly optimizing beamforming, user association, and UAV
trajectories under heterogeneous QoS and sensing constraints. To extend
the single ISAC node scenario to ISAC networks, \cite{UAV_ISAC_6}
investigates multi-BS, multi-UAV ISAC networks by jointly optimizing
transmit beamforming, user association, and the 3D UAV trajectories.
This coordinated design satisfies the heterogeneous sensing and communication
requirements of multiple UAVs under power constraints, thereby achieving
a substantial improvement in resource utilization efficiency. 

\textbf{UAV-aided ISAC AGICNs.} Beyond acting as sensing targets or
communication nodes, UAVs are emerging as active and indispensable
components of AGICNs. Thanks to UAVs' controllable mobility, favorable
aerial visibility, and flexible deployment, they can enhance both
communication and sensing capability in complex or extreme environments,
particularly when natural disasters render terrestrial infrastructures
inoperative and create “information islands.” In such cases, large-scale
failures of ground nodes leave users unable to access to BSs. To address
this issue, \cite{ISAC_UAV_new_1} and \cite{ISAC_UAV_new_2} propose
equipping UAV platforms with passive reflective RISs, enabling them
to operate as lightweight, energy-efficient aerial relays that redirect
ISAC beams from the BS toward users in disaster-stricken regions,
thereby achieving ultra–low-power emergency coverage. Further, \cite{UAV_com_new_4}
develops a joint optimization framework based on DDPG and game theory
to determine transmit power, RIS reflection coefficients, and RIS
activation ratios, delivering substantial EE gains over baseline methods.

\textbf{ISAC-UAV in AGICNs.} By integrating multi-modal sensors (e.g.,
force, vision, and LiDAR) and leveraging their mobility and wide field
of view, UAVs can acquire both self-state and environmental information
to support diverse sensing and communication tasks \cite{UAV_sensing_1,UAV_sensing_3,UAV_sensing_4,UAV_sensing_2}.
From a sustainability perspective, ISAC-UAVs can be used to monitor
soil moisture, glacier melt, and urban heat island effects, which
provides critical data for building large-scale models of carbon emissions
and climate change. Study \cite{UAV_ISAC_3} constructs an ISAC-enabled
multi-UAV network in which UAVs evolve from traditional communication
relays into fully active dual-functional nodes capable of both communication
and sensing. By jointly optimizing UAV trajectories, user association,
and beamforming, the UAVs can flexibly adjust their spatial positions
and radiation patterns, thereby simultaneously enhancing high-quality
communication services and high-resolution sensing for ground users.

This subsection reviewed ISAC-oriented techniques that enable holistic
sensing in AGICNs, including ISAC beamforming design, UAV-assisted
sensing and communication, and multi-UAV collaborative sensing frameworks.
By integrating sensing and communication functionalities, these approaches
exploit shared spectrum, hardware, and spatial degrees of freedom
to improve both environmental perception and communication efficiency.
A key insight is that ISAC-based designs significantly enhance system
awareness and resource utilization compared with conventional separated
architectures, particularly in dynamic and data-intensive scenarios.

\subsection{Wireless Energy and Electronic Environments Coordinated Design in
AGICNs}

Although UAVs' high mobility, flexible deployment, and ease of establishing
LoS communication links enhance AGICNs' coverage capability and capacity
compared to conventional ground networks, the limited battery capacity
and high energy consumption of UAVs remain major bottlenecks for their
large-scale and long-duration operations \cite{EH_3,EH_2,EH_1}. Traditional
ground-based recharging introduces substantial time costs and operational
expenses, severely restricting the scalable development of the low-altitude
economy \cite{EH_4}. Consequently, how to extend the life of UAVs
while ensuring the network performance has become a key research direction.
WEH technology has emerged as a promising sustainable power supply
approach \cite{Technology_1}. By extracting energy from the surrounding
environment, including RF signals, solar power, and wind energy, WEH
enables UAVs to achieve self-sustained operation, thereby prolonging
their service lifetime and reducing dependence on conventional batteries
\cite{EH_7,EH_8}. This makes WEH a key research focus for building
greener and more resilient AGICNs. In this work, we compare four major
WEH technologies, i.e., single UAV SWIPT, multi-UAVs SWIPT, RIS-aided
UAV SWIPT, and simultaneous transmitting and reflecting-reconfigurable
intelligent surfaces (STAR-RIS). 
\begin{figure*}[t]
\centering
\begin{centering}
\includegraphics[width=17.6cm]{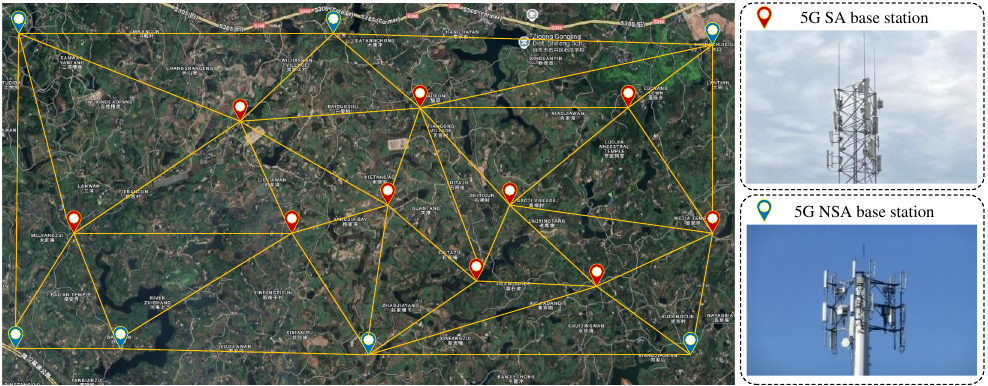}
\par\end{centering}
\caption{The terrestrial communication networks topology and the results of
triangulation for G2A coverage in Zigong, Sichuan province, China.
\label{fig:Ground-to-air-coverage-testbed}}
\end{figure*}

\textbf{Single UAV SWIPT.} To enable UAVs to harvest energy from RF
signals, the concept of SWIPT was first introduced by \cite{SWIPT_1,SWIPT_2}.
SWIPT allows RF signals to simultaneously carry both information and
energy, enabling UAVs not only to communicate with ground devices
but also to wirelessly charge them, or alternatively, to replenish
their own energy by harvesting signals transmitted from the ground
\cite{SWIPT_3,SWIPT_4}. Study \cite{SWIPT_17} proposes a UAV-assisted
IoT SWIPT system to provide sustainable energy supply and communication
services for IoT devices in remote or disaster-stricken regions. In
the system, a power-splitting receiver model is adopted to characterize
the trade-off between information transmission and energy transfer.
A G2A channel model dominated by LoS links is introduced to accurately
capture the aerial transmission characteristics between UAVs and IoT
devices. Similar research \cite{SWIPT_16,SWIPT_18} both focus on
jointly optimizing the UAV trajectory and power allocation strategy
under constraints of UAV transmit power and flight speed, with the
goal of maximizing the minimum harvested energy among all IoT devices
during a mission period, while simultaneously ensuring that the average
data rate requirement of each device is satisfied. 

\textbf{Multi-UAVs SWIPT.} To enhance coverage capacity, study \cite{SWIPT_18}
extends single-UAV SWIPT to a multi-UAV-assisted IoT system. The framework
jointly optimizes UAV deployment, user association, and power allocation
to balance data rate and wireless energy harvesting requirements.
An iterative algorithm combining alternating optimization (AO) and
successive convex approximation (SCA) is proposed. This algorithm
decomposes the original optimization problem into two interdependent
subproblems, i.e., user association optimization and joint optimization
of UAV 3D positions and power allocation. The former is solved as
an integer programming problem given fixed UAV positions and power
allocation. And the latter is inherently non-convex and tackled by
converting it into a convex form via the SCA technique. By alternately
solving these two subproblems, the algorithm iteratively refines the
solution until convergence, effectively balancing network throughput,
fairness, and energy sustainability in multi-UAV-assisted SWIPT IoT
systems. 

\textbf{RIS-aided UAV SWIPT.} To address the LoS degradation problem
in complex environments, studies \cite{SWIPT_RIS_1,SWIPT_RIS_2} introduce
RISs into SWIPT-enabled AGICNs. The deployment of RISs aims to manipulate
the wireless environment by intelligently controlling the electromagnetic
field and wave propagation, thereby enhancing WEH efficiency and SE
for both ground devices and UAVs. Specifically, study \cite{SWIPT_RIS_1}
investigates the joint optimization of UAV trajectory, transmit precoding,
and RIS phase shifts to maximize the minimum information rate among
all information receivers, while considering practical constraints
including no-fly zones for UAVs and RIS phase dispersion. Furthermore,
\cite{SWIPT_RIS_2} introduces non-orthogonal multiple access (NOMA)
into the joint communication and energy transfer scenario within AGICNs.
The authors study how to jointly optimize the UAV trajectory, successive
interference cancellation decoding order in NOMA, UAV transmit power
allocation, PS ratios, and IRS reflection coefficients to maximize
the achievable system sum rate.

\textbf{STAR-RIS in AGICNs.} To further improve UAV WEH performance,
study \cite{SWIPT_RIS_3} considers mounting RISs directly on UAVs
to enlarge the wireless energy reception area. The framework jointly
optimizes UAV deployment, user scheduling, and RIS configuration to
improve energy efficiency under communication and WEH constraints.
Additionally, leveraging the transmission capability of RISs, \cite{SWIPT_RIS_4}
explores a STAR-RIS-assisted UAV communication system to overcome
the limitation that conventional RISs can only cover its one side.
In this scenario, wireless signals are simultaneously reflected and
transmitted, thereby serving users on both sides of the RIS. The problem
is formulated as the joint optimization of STAR-RIS reflection and
transmission beamforming vectors, 3D UAV trajectory, and power allocation,
with the objective of maximizing the sum rate of all users. The results
demonstrate that STAR-RIS achieves a higher total system throughput
than conventional RISs and that the UAV’s optimal trajectory tends
to move closer to the STAR-RIS deployment location. 
\begin{table*}[t]
\caption{The beam configuration of G2A coverage and its application. \label{tab:The-beam-configuration}}

\centering{}%
\begin{tabular}{|>{\centering}m{1.2cm}|>{\centering}m{2cm}|>{\centering}m{2cm}|>{\centering}m{2.5cm}|c|}
\hline 
{\footnotesize\textbf{Scenario ID}} & {\footnotesize\textbf{Horizontal 3dB beam width $\phi_{H}$}} & {\footnotesize\textbf{Vertical 3dB beam width $\phi_{V}$}} & {\footnotesize\textbf{Available inclination angle}} & {\footnotesize\textbf{Applications}}\tabularnewline
\hline 
{\scriptsize 17} & {\scriptsize 60$\lyxmathsym{\textdegree}$} & {\scriptsize 6$\lyxmathsym{\textdegree}$} & {\scriptsize -20$\lyxmathsym{\textdegree}$\textasciitilde -10$\lyxmathsym{\textdegree}$} & {\scriptsize High-altitude (> 300m), strong direct beam.}\tabularnewline
\hline 
{\scriptsize 18} & {\scriptsize 120$\lyxmathsym{\textdegree}$} & {\scriptsize 6$\lyxmathsym{\textdegree}$} & {\scriptsize -9$\lyxmathsym{\textdegree}$\textasciitilde -5$\lyxmathsym{\textdegree}$} & {\scriptsize High-altitude (> 300m), wide horizontal beam.}\tabularnewline
\hline 
{\scriptsize 19} & {\scriptsize 60$\lyxmathsym{\textdegree}$} & {\scriptsize 15$\lyxmathsym{\textdegree}$} & {\scriptsize -20$\lyxmathsym{\textdegree}$\textasciitilde -10$\lyxmathsym{\textdegree}$} & {\scriptsize Middle/low altitude (100\textasciitilde 300m), strong
direct beam.}\tabularnewline
\hline 
{\scriptsize 20} & {\scriptsize 120$\lyxmathsym{\textdegree}$} & {\scriptsize 15$\lyxmathsym{\textdegree}$} & {\scriptsize -9$\lyxmathsym{\textdegree}$\textasciitilde -5$\lyxmathsym{\textdegree}$} & {\scriptsize Middle/low altitude (100\textasciitilde 300m), wide horizontal
beam}\tabularnewline
\hline 
\end{tabular}
\end{table*}

This subsection reviewed techniques for wireless energy provisioning
and electromagnetic environment coordination in AGICNs, including
multi-UAV SWIPT systems, UAV-mounted RIS/STAR-RIS architectures, and
energy-aware trajectory and resource optimization. These approaches
aim to jointly manage energy transfer, communication, and propagation
environments to enhance coverage efficiency and reduce overall energy
consumption. A key insight is that integrating wireless power transfer
with adaptive propagation control (e.g., RIS-enabled environments)
can significantly improve energy utilization and extend network lifetime,
especially in energy-constrained aerial scenarios.

\section{Testbed and Experimental Validation for Low-Carbon-oriented AGICNs}

In Sections III and IV, we introduced an ISCC-driven architectural
framework for sustainable AGICNs and systematically reviewed its key
enabling technologies, including AI, low-carbon communication, ISAC,
and WEH. To further assess the feasibility and practical effectiveness
of this architecture, we perform a series of experiments under real
G2A propagation conditions, hardware limitations, and complex terrain. 

\subsection{Testbed Deployment and System Configuration}

A real-world AGICN coverage test platform has been deployed in Zigong,
Sichuan province, China. This platform enables empirical evaluation
of AGICNs performance and supports enhancements in ISCC-driven coverage
capability and carbon-reduction. The subsequent sections describe
the system configuration, experimental methodology, and major experiment
results. 

\textbf{Physical environment selection.} Zigong is selected as the
test environment because it offers diverse terrain and channel characteristics,
including medium-density urban areas, rivers, hilly regions, and open
suburban zones. These features provide a rich set of scenarios for
evaluating different G2A channel models. The test platform spans approximately
44 square kilometers, extending about 11 km from east to west and
4 km from north to south, as illustrated in Fig. \ref{fig:Ground-to-air-coverage-testbed}. 

\textbf{Ground communication networks and edge computation devices
configuration.} Within this area, ten 5G standalone (SA) BSs and seven
non-standalone (NSA) public BSs have been deployed, with inter-site
distances ranging from 1.5 km to 2.5 km. These SA BSs operate in the
3.5 GHz C-band and are equipped with 64T$\times$64R massive MIMO
antenna arrays that support high-precision 3D beamforming and remote
electrical tilt (RET). Three computation servers, including two Dell
R750 XS edge services and one Dell R750 edge service, are deployed
to serve as edge computation nodes and to host the ISCC computation
subsystem. These nodes process sensing data from UAVs, execute AI-based
optimization algorithms, and store the associated AI models. 

\textbf{Aerial mobile ISCC platforms configuration. }The experiments
employ several multirotor UAVs equipped with (i) a 5G terminal module
for connecting to the terrestrial SA network, transmitting telemetry
data, and receiving control signaling, and (ii) high-precision sensors,
including real-time kinematic (RTK)-GPS for accurate positioning,
an inertial measurement unit, and a LiDAR unit for environmental sensing
and obstacle avoidance. During the G2A coverage tests, the UAVs follow
predefined trajectories at altitudes of 100 m (standard coverage),
200 m (regional coverage), and 300 m (wide-area coverage). Specifically,
100 m corresponds to dense urban scenarios with strong blockage effects,
200 m reflects a transitional regime with improved LoS conditions,
and 300 m represents higher-altitude operations with more stable propagation
characteristics \cite{Testbed_1}.

\textbf{The structure and function deployment of AI model. }In our
work, a DRL–based AI agent is deployed on the edge computation server
to generate energy-efficient G2A coverage strategies. The agent is
built upon a hierarchical neural network architecture designed to
address the optimization of a high-dimensional hybrid action space.
Functionally, the architecture is divided into two cascaded components.
The neural layers near the input act as a classifier, selecting the
optimal initial pattern for each BS from the four standard 3GPP high-altitude
beam modes (as shown in Tab. \ref{tab:The-beam-configuration}). The
layers near the output operate as a regressor, performing fine-grained
continuous control based on the selected pattern, which adjusts the
RET, horizontal, and vertical beamwidth within a \textpm 3° range.
This hierarchical design enables the AI agent to jointly handle discrete
beam-pattern selection and continuous parameter optimization, thereby
supporting efficient G2A coverage adaptation. 

\begin{figure}[t]
\centering
\begin{centering}
\includegraphics[width=8cm]{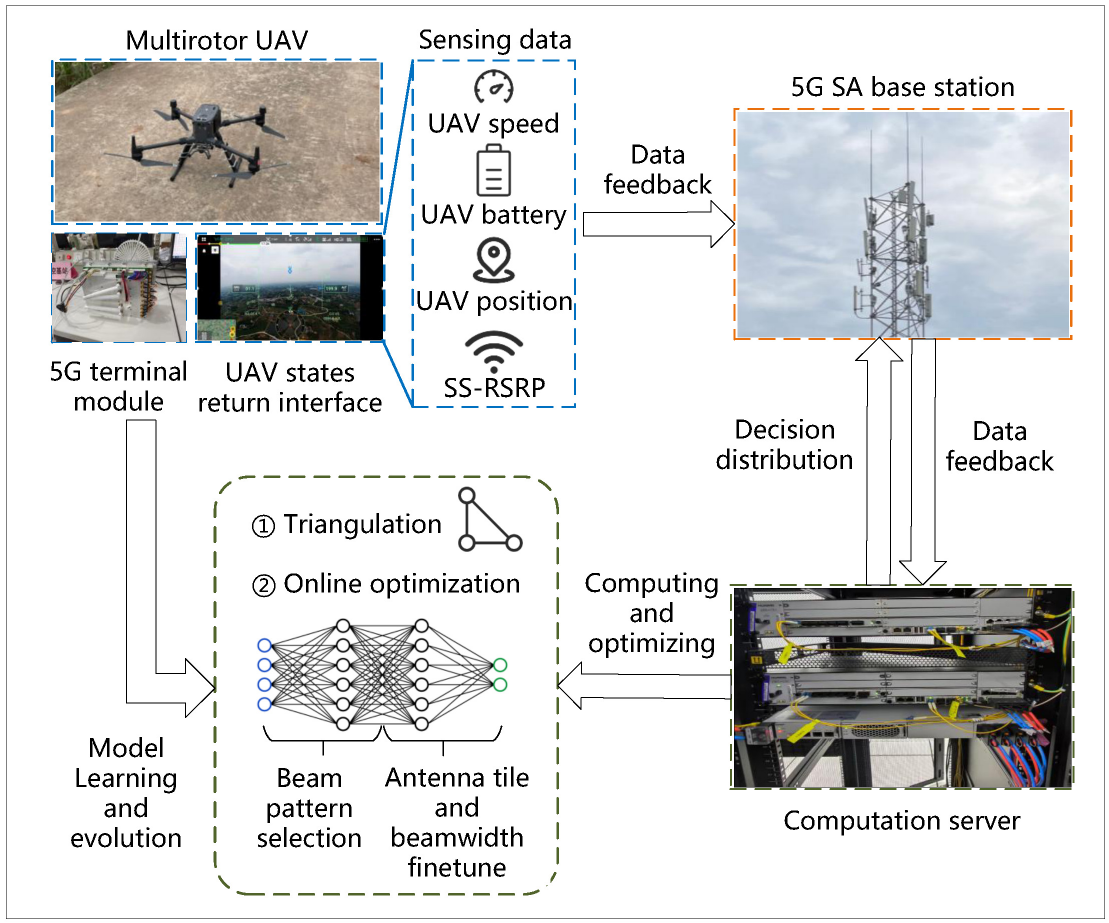}
\par\end{centering}
\caption{The illustration of proposed energy-efficient G2A coverage enhancement
method. \label{fig:Energy-efficient-G2A-coverage}}
\end{figure}

\textbf{Multimodal sensing states fusion for AI model input representation.}
Considering the heterogeneity and multimodal characterizes of the
sensed information, which includes UAV state variables (e.g., UAV
position, velocity vector, battery level, RSRP, and SINR) and network-side
parameters (e.g., topology, antenna configuration, and inter-cell
interference levels), the mobile edge computation (MEC) server first
performs spatiotemporal alignment and normalization across all data
sources. The resulting preprocessed feature vector is then fed into
the input layer of the DRL network, enabling the agent to construct
an implicit representation of coverage holes and interference distributions
for the current time slot. This latent representation provides a reliable
basis for subsequent beamforming decisions.

\textbf{ISCC-based closed-loop continue learning and updating.} To
enhance the generalization of this AI-based G2A coverage method, we
design a reward mechanism based on maximizing the spatial signal coverage
probability (SSCP). During online inference, the AI agent outputs
action vectors to adjust beams and perform CoMP transmission. The
system computes rewards from measured feedback. A positive reward
is obtained only if SSCP exceeds 90\% within the target airspace and
transmit power meets energy constraints. Otherwise, a penalty is applied.
Through this trial-and-feedback learning process, the network updates
its weights via backpropagation, learning optimal strategies that
transform inter-cell interference into useful signals.

\subsection{Energy-efficient G2A coverage enhancement method}

To address the high energy consumption and severe interference challenges
inherent in G2A coverage, an ISCC-based energy-efficient G2A coverage
enhancement method is proposed. The objective is to maximize the SSCP
with given UAV flight trajectories and transmit power constraints
through jointly optimizing the antenna tilt, horizontal and vertical
beamwidths of G2A beams. The detailed procedures are described as
follows (as shown in Fig. \ref{fig:Energy-efficient-G2A-coverage}).

\textbf{Initialize}: Each SA 5G BS randomly selects one of the four
3GPP-defined G2A beam patterns to provide 3D coverage for UAVs. UAVs
associate with the BS which offers themselves the maximum reference
signal received power (RSRP). 

\textbf{Step 1}: Sensing and data feedback. The UAV periodically senses
its 3D position via RTK-GPS, battery status, flight speed, and the
received SINR and RSRP maps. It then reports these measurements to
the serving BS. These measurements are transmitted to the MEC servers
through the 5G SA BS.

\textbf{Step 2}: Computing and optimizing. The edge servers input
the communication network topology, UAV locations, and PSPR measurements,
and output optimized beam and antenna configurations for the BSs.
The computation consists of following main components.
\begin{itemize}
\item First, a multi-BSs cooperative triangulation method is performed,
which leverages coordinated multi-point transmission to enable multiple
ground BSs to jointly serve UAV users. In this way, the inter-cell
interference can be converted into useful signals. 
\item To achieve energy-efficient coverage, we aim to maximize the SSCP,
defined as the ratio between the accumulation of aerial space in which
the received signal strength exceeds a predefined threshold $\gamma_{\mathrm{th}}$
and the accumulation of target 3D coverage space $\mathcal{S}$. The
optimization problem is formulated as following:

\begin{align}
(\mathrm{P1}):\:\:\mathrm{\underset{\{\theta_{m}\}}{max}}\:\: & \mathrm{SSCP}(\{\theta_{m}\})\nonumber \\
 & =\frac{\int_{\mathcal{S}}\boldsymbol{1}(P_{r}(x,y,z;\{\theta_{m}\})\geq\gamma_{\mathrm{th}})dxdydz}{\mathcal{S}}\nonumber \\
\mathrm{s.t.}\:\: & \theta_{\mathrm{min}}\leq\theta_{m}\leq\theta_{\mathrm{max}},\;\forall m\in\mathcal{M},
\end{align}

where $\{\theta_{m}\}$ is the tilt angle of BS antennas. $\theta_{\mathrm{min}}$
and $\theta_{\mathrm{max}}$ denote the minimum and maximum allowable
tilt angles, respectively. $\boldsymbol{1}(\cdot)$ is an indicator
function that equals 1 when the RSRP value $P_{r}$ at location $(x,y,z)$
meets the threshold $\gamma_{\mathrm{th}}$, and 0 otherwise. 
\item Then, considering that the demand for radiated energy in the vertical
direction significantly increases while the horizontal direction is
reduced, due to the non-negligible altitudes of UAVs and the application
of convex triangulation among BSs. It becomes necessary to design
new G2A beams that can accommodate the increased vertical coverage
demand while maintaining EE. To overcome this issue, we incorporate
horizontal and vertical beam constraints into P1, considering the
asymmetric gain distribution of antenna beams in the horizontal and
vertical planes. The constraints are given by

\begin{align}
\phi_{H}^{\mathrm{min}}\leq\phi_{H}\leq\phi_{H}^{\mathrm{max}},\: & \phi_{V}^{\mathrm{min}}\leq\phi_{V}\leq\phi_{V}^{\mathrm{max}},\\
\phi_{V} & =\phi_{H}^{\alpha_{V}/\alpha_{H}},\label{eq:asymmetric gain}
\end{align}

where $\phi_{H}$ denotes the horizontal beam spread, $\phi_{V}$
denotes the vertical beam spread. The constraint (\ref{eq:asymmetric gain})
denotes the asymmetry of beam gain in horizontal and vertical directions.
$\alpha_{V}$ and $\alpha_{H}$ represent the signal propagation path
loss factors in horizontal and vertical directions, respectively.
The received signal strength $P_{r}$ at location $(x,y,z)$ can be
calculated by 

\begin{align}
P_{r}(x,y,z; & \{\theta_{m}\};\phi_{H},\phi_{V})\nonumber \\
= & P_{t}G(\{\theta_{m}\};\phi_{H},\phi_{V})L^{-1}(d).
\end{align}

Here, $P_{t}$ is transmit power, $G(\{\theta_{m}\};\phi_{H},\phi_{V})$
is a function of both beam shape $\phi_{H},\phi_{V}$ and elevation
$\{\theta_{m}\}$, $L^{-1}(d)$ is the path loss as a function of
distance $d$. 
\end{itemize}

\begin{itemize}
\item The DRL-based AI agent deployed on the server then optimizes the decision
variables according to the defined objective and the current state
of the AGICNs. The agent receives a positive reward when the SSCP
exceeds 90\% and a penalty otherwise. 
\end{itemize}
\quad \textbf{Step 3}: Decision distribution and actuation. After
the AI agent generates its decision, the optimized decision is delivered
to the BSs through the core network. Each BS then updates strategy
to serve the UAV and continues collecting measurement reports, which
are sent back to the computation subsystem for further processing.
\begin{figure}[t]
\begin{centering}
\includegraphics[width=9cm]{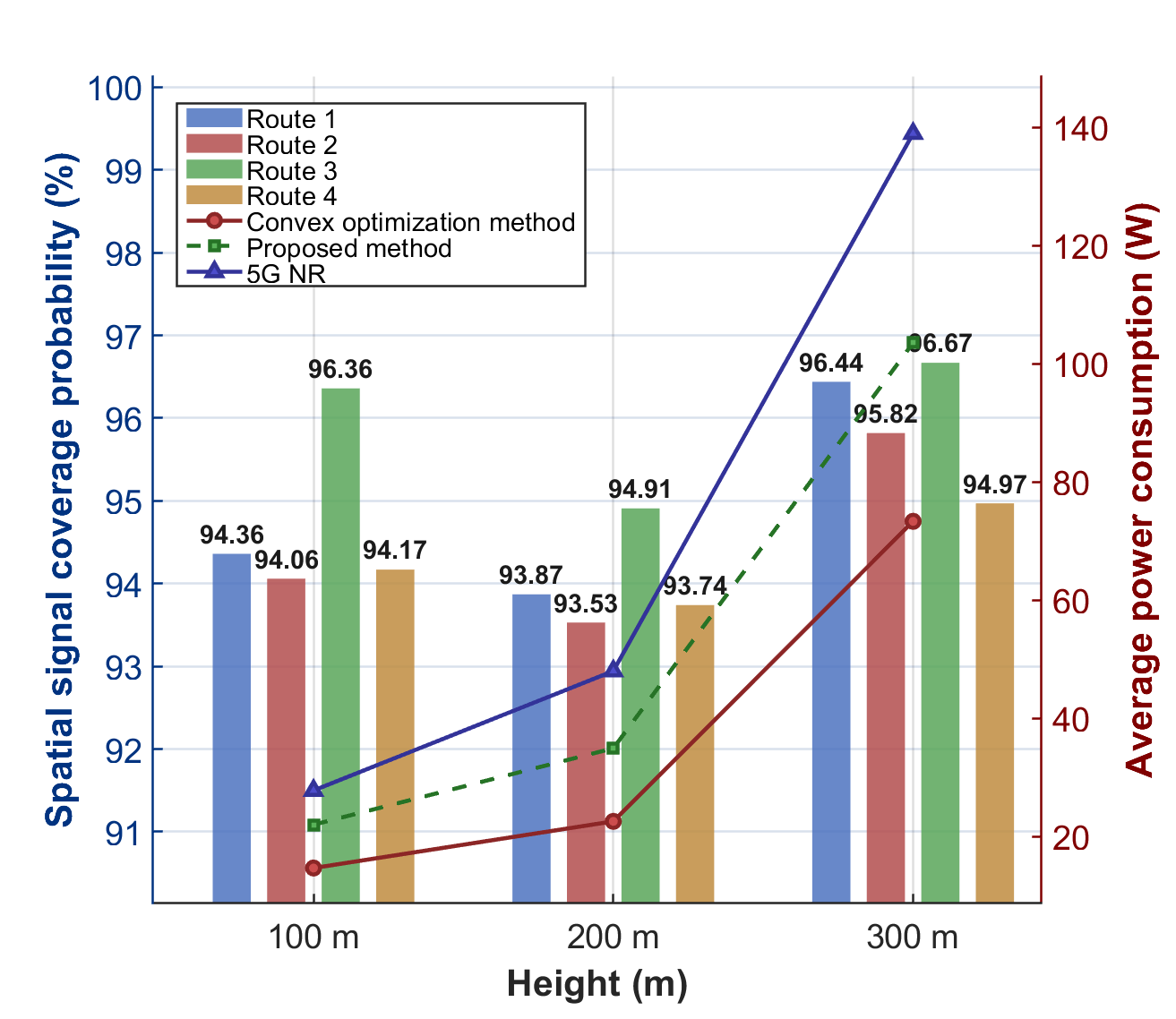}
\par\end{centering}
\caption{Experimental results of G2A coverage. \label{fig:Ground-to-air-experimental-resul}}

\vspace{-0.5cm} 
\end{figure}

It is worth noting that there exists an inherent trade-off between
coverage performance and energy consumption. For instance, increasing
UAV altitude or transmit power can improve coverage by enhancing LoS
conditions, but may also lead to higher propulsion or transmission
energy consumption. In this work, the proposed ISCC framework addresses
this trade-off through closed-loop optimization, where both coverage
performance and energy consumption are jointly considered. We formulate
energy minimization as the optimization objective while enforcing
UAV's RSRP as a constraint. This allows the DRL agent to learn adaptive
resource allocation strategies under dynamic conditions. As a result,
the system seeks a balanced operating point that achieves high coverage
while maintaining energy-efficient operation.

\begin{table*}[t]
\begin{centering}
\caption{Sampling Statistics and Coverage Performance for Each Route. \label{tab:Sampling-Statistics-and}}
\par\end{centering}
\begin{onehalfspace}
\begin{centering}
\begin{tabular}{|c|>{\centering}m{2cm}|>{\centering}m{2.5cm}|c|>{\centering}m{2cm}|>{\centering}m{2.5cm}|c|}
\hline 
\multirow{2}{*}{{\scriptsize\textbf{Flight Route}}} & \multicolumn{3}{c|}{{\scriptsize\textbf{Route 1}}} & \multicolumn{3}{c|}{{\scriptsize\textbf{Route 2}}}\tabularnewline
\cline{2-7}
 & {\scriptsize Total Sampling Points} & {\scriptsize Points with RSRP \ensuremath{\ge} \textminus 90 dBm} & {\scriptsize SSCP (\%)} & {\scriptsize Total Sampling Points} & {\scriptsize Points with RSRP \ensuremath{\ge} \textminus 90 dBm} & {\scriptsize SSCP (\%)}\tabularnewline
\hline 
{\scriptsize\textbf{100 m}} & {\scriptsize 7881} & {\scriptsize 7212} & {\scriptsize 91.51\%} & {\scriptsize 8654} & {\scriptsize 7965} & {\scriptsize 92.04\%}\tabularnewline
\hline 
{\scriptsize\textbf{200 m}} & {\scriptsize 6225} & {\scriptsize 5898} & {\scriptsize 94.75\%} & {\scriptsize 6693} & {\scriptsize 6120} & {\scriptsize 91.44\%}\tabularnewline
\hline 
{\scriptsize\textbf{300 m}} & {\scriptsize 7581} & {\scriptsize 7032} & {\scriptsize 92.76\%} & {\scriptsize 6705} & {\scriptsize 6501} & {\scriptsize 96.96\%}\tabularnewline
\hline 
\multirow{2}{*}{{\scriptsize\textbf{Flight Route}}} & \multicolumn{3}{c|}{{\scriptsize\textbf{Route 3}}} & \multicolumn{3}{c|}{{\scriptsize\textbf{Route 4}}}\tabularnewline
\cline{2-7}
 & {\scriptsize Total Sampling Points} & {\scriptsize Points with RSRP \ensuremath{\ge} \textminus 90 dBm} & {\scriptsize SSCP (\%)} & {\scriptsize Total Sampling Points} & {\scriptsize Points with RSRP \ensuremath{\ge} \textminus 90 dBm} & {\scriptsize SSCP (\%)}\tabularnewline
\hline 
{\scriptsize\textbf{100 m}} & {\scriptsize 8448} & {\scriptsize 8298} & {\scriptsize 98.22\%} & {\scriptsize 8723} & {\scriptsize 8509} & {\scriptsize 97.55\%}\tabularnewline
\hline 
{\scriptsize\textbf{200 m}} & {\scriptsize 6991} & {\scriptsize 6707} & {\scriptsize 95.94\%} & {\scriptsize 7975} & {\scriptsize 7381} & {\scriptsize 92.55\%}\tabularnewline
\hline 
{\scriptsize\textbf{300 m}} & {\scriptsize 7944} & {\scriptsize 7838} & {\scriptsize 98.67\%} & {\scriptsize 7941} & {\scriptsize 7601} & {\scriptsize 95.72\%}\tabularnewline
\hline 
\end{tabular}
\par\end{centering}
\end{onehalfspace}
\vspace{-0.5cm} 
\end{table*}

\subsection{Experiment results and analysis}

The statistical results of three flight altitudes (100 m, 200 m, and
300 m) and four predefined flight routes at each altitude are summarized
in Tab. \ref{tab:Sampling-Statistics-and}. Signal samples are collected
at a fixed rate of 5 Hz. With an average UAV flight speed of 15 m/s,
the spatial sampling interval is approximately 3 m. In practical measurements,
SSCP is approximated as the ratio of the number of sampled points
satisfying the threshold to the total number of measurement points,
expressed as 

\begin{equation}
\mathrm{SSCP}=\frac{\text{Number of points with }\mathrm{RSRP}\ge-90\,\mathrm{dBm}}{\text{Total number of measurement points}}.
\end{equation}
In this study, the RSRP threshold is set to \textminus 90 dBm, which
corresponds to the minimum coverage requirement commonly adopted in
practical cellular network deployment. To ensure statistical robustness,
each flight route at each altitude is measured five times, and the
reported SSCP values are obtained by averaging over all measurements.
The test results show that the proposed ISCC-based energy-efficient
G2A coverage method enables UAVs to maintain an SSCP above 90\% across
three flight altitudes and four flight routes. 

For energy consumption evaluation, we adopt a conventional 5G NR–based
G2A coverage configuration as the baseline. Specifically, the baseline
operates with a set of predefined beam configurations, as listed in
Tab. \ref{tab:The-beam-configuration}. For each UAV trajectory segment,
a fixed beam pattern is selected from these configurations, without
dynamic adaptation or mechanical downtilt adjustment. In contrast
to the proposed ISCC-based method, which enables fine-grained beam
control and adaptive downtilt tuning, the baseline method relies on
static beam selection and does not perform real-time optimization.
Compared with this 5G NR baseline, the proposed ISCC-based method
achieves power consumption reductions of 20.8\%, 27.25\%, and 25.48\%
at flight altitudes of 100 m, 200 m, and 300 m, respectively, while
maintaining comparable coverage performance, as shown in Fig. \ref{fig:Ground-to-air-experimental-resul}. 

These improvements stem from the integrated effects of sensing, optimization,
and cooperation in the ISCC-based design. By incorporating real-time
UAV state awareness and MEC-assisted intelligent decision making,
the system dynamically adjusts downtilt angles, horizontal/vertical
beamwidths, and high-altitude beam modes based on the UAV’s position,
interference distribution, and channel quality. This mechanism ensures
that the UAV remains within the main beam or cooperative main beam,
thereby avoiding the coverage degradation in fixed-beam configurations
under altitude variations. The triangulation-based cooperative transmission
converts multi-cell LoS interference into diversity gains, which improves
desired signal significantly. Furthermore, aided by AI-enabled joint
optimization of beam configuration, cooperative transmission cluster,
and transmit power allocation, the radiated energy is concentrated
exclusively toward spatial directions that contribute meaningfully
to G2A coverage. This directional energy distribution substantially
suppresses redundant emissions while maintaining the required coverage
performance.

It is worth noting that the testbed is deployed in Zigong, which includes
diverse terrains such as urban, suburban, and hilly areas, representing
typical low-altitude operation scenarios. While G2A channel characteristics
may vary across different environments due to factors such as building
density and terrain variations, the proposed ISCC framework is not
restricted to a specific terrain type. By leveraging environment-aware
sensing and data-driven decision-making, the system can adapt to varying
propagation conditions through closed-loop optimization. Therefore,
although the exact numerical results may differ under different terrains,
the overall performance trends, including high coverage probability
and energy savings, remain consistent.

\subsection{Scalability Analysis for Large-Scale AGICNs}

In Section V-C, we have demonstrated the improvement of the ISCC control
loop in the sustainability of AGICNs. Simulations show significant
results in a small-scale network (17 BSs), achieving over 90\% SSCP
and a 20\% to 27\% power reduction. However, as network size grows,
interference and algorithm scalability become critical. In this subsection,
we analyze the scalability of our framework. 

\textbf{Interference}. With $N$ BSs and $M$ UAVs, the number of
interfering links scales as $M\times N$. When $M=\Theta(N)$, this
becomes $\mathcal{O}(N^{2})$, which is prohibitive. Delaunay triangulation
groups neighboring BSs into cooperative triplets. Through CoMP, interference
is converted into useful signals, reducing the effective interference
links to $\mathcal{O}(N)$. Since the triangulation itself costs $\mathcal{O}(N\mathrm{log}N)$,
the overall management complexity becomes near-linear.

\textbf{Coordination}. Centralized power optimization requires each
BS to coordinate with all others, leading to a complexity that scales
as $\mathcal{O}(N^{2})$. The ISCC framework adopts a multi\nobreakdash-agent
and hierarchical architecture (Section IV-A, Fig.\,\ref{fig:ISCC-enabled-multi-UAV-control}).
The network is partitioned into $K\ll N$ independent clusters, each
with a dedicated DRL agent that coordinates only with neighbors. This
reduces management complexity from $\mathcal{O}(N^{2})$ to $\mathcal{O}(K^{2})$,
significantly improving scalability.

In summary, while ISCC demonstrates substantial gains in coverage
and EE at small network scales, its practical deployment in large-scale
AGICNs requires careful handling of interference growth and computational
complexity. By leveraging structured coordination mechanisms such
as Delaunay triangulation and adopting a distributed, multi-agent
hierarchical architecture, the framework effectively transforms quadratic
scaling behaviors into near-linear ones. This enables ISCC to maintain
manageable complexity while preserving its performance advantages,
thereby providing a viable pathway toward scalable, sustainable low-altitude
network operation.

\section{Open Research Issues and Future Directions}

Despite significant progress in improving the energy efficiency (EE)
and sustainability of AGICNs, the realization of a truly sustainable
and intelligent system remains challenged by multiple technical bottlenecks
spanning communication, sensing, computation, and energy. Addressing
these multidimensional issues is essential for enabling scalable,
resilient, and carbon-efficient AGICNs. To this end, the following
open research directions are discussed in relation to different stages
of the ISCC closed-loop, including sensing, communication, computation,
execution, and learning.
\begin{itemize}
\item \textbf{High dynamics and heterogeneous integration \cite{Future_1,History_4}.}
Terrestrial and aerial networks differ significantly in link type
(LoS vs. NLoS), mobility, and coverage scale, making a unified protocol
stack an open challenge. Achieving millisecond-level latency and ultra-high
reliability (e.g., 10\textsuperscript{-}\textsuperscript{9}) for mission-critical
low-altitude applications requires real-time resource reservation
in dynamic topologies, which current architectures cannot support.
These challenges primarily affect the sensing and communication stages
of the ISCC closed-loop, where rapid dynamics degrade perception accuracy
and link stability. Such uncertainties propagate to the computation
stage, increasing the complexity of real-time resource optimization
and decision-making.
\item \textbf{Generalization of AI models in AGICNs \cite{Future_2,Future_3}.}
AI-based scheduling in AGICNs faces generalization challenges due
to heterogeneous environments (varying region, altitude, task type,
climate), causing distribution shifts between training and deployment.
Current models degrade in decision accuracy under cross-domain variability.
Developing algorithms that generalize across dynamic aerial environments
while meeting strict latency and energy constraints remains a key
barrier. This issue mainly affects the computation and learning stages
in the ISCC closed-loop, where robust models are needed for resource
orchestration and continuous adaptation to evolving conditions.
\item \textbf{Multi-modal fusion and ISAC co-design \cite{Future_4,Future_6,Future_5}.}
Data from terrestrial and aerial sensors (optical, radar, and RF)
vary in sampling rates, resolution, and timestamps. Minimizing synchronization
errors and achieving high-precision spatio-temporal fusion are essential
for training reliable AI agents of the AGICNs. Furthermore, realizing
ISAC functionality on shared hardware and spectrum remains a major
technical barrier. Communication systems prioritize throughput and
spectral efficiency, whereas sensing focuses on detection accuracy
and spatial resolution. Reconciling these fundamentally divergent
design objectives demands novel waveform, coding, and signal processing
co-design. 
\item \textbf{Space-air-ground integration networks and global sustainability
\cite{space-air-ground-2,space–air–ground-1}.} As 6G advances
toward space–air–ground integration, incorporating satellites into
AGICNs is emerging as a new paradigm for achieving sustainable development.
Satellite networks, with their wide-area coverage capability, provide
resilient backhaul links for AGICNs and supply global remote-sensing
data to ISCC-based collaborative control loops, thereby enriching
the diversity of training samples for AI agents. Leveraging satellite
computing, energy-intensive and non-real-time tasks can be offloaded
to low-earth-orbit satellites, reducing the power consumption of aerial
and terrestrial nodes. Moreover, the solar energy harvested by satellites
can be transmitted via microwave or laser to HAPs, further lowering
the carbon emissions of AGICNs. 
\end{itemize}

\section{Conclusion}

This work establishes a sustainable AGICN framework for 6G and low-altitude
economies, shifting from coverage-centric to carbon-aware designs.
Key contributions include: (1) an ISCC-driven closed-loop control
framework integrating communication, sensing, and computation for
dynamic carbon optimization; (2) sustainability metrics including
carbon intensity and full lifecycle assessment; and (3) experimental
validation achieving >90\% coverage with 20\% power reduction. The
study analyzes AI-empowered beam prediction, hybrid precoding, and
SWIPT for energy-carbon reduction. Zigong testbed results confirm
practical viability. Future work includes heterogeneous integration,
AI generalization, cross-domain data fusion, and space-air-ground
integration. This work provides theoretical and practical guidance
for carbon-neutral aerial-ground networks and sustainable low-altitude
infrastructure.

\bibliographystyle{ieeetr}
\bibliography{Reference_survey}

\end{document}